\documentclass[preprint2,times,tighten]{aastex6}
\usepackage{amsmath,amstext}
\usepackage{newtxmath} 
\usepackage[T1]{fontenc}
\usepackage{url}
\usepackage{ulem}

\newcommand{\cmpc}{$h$\,cMpc$^{-1}$}
\newcommand{\exn}{{excess noise}}
\newcommand{\exv}{{excess variance}}
\newcommand{\ddcal}{{\sl DD}-calibration}
\newcommand{\dical}{{\sl DI}-calibration}
\newcommand{\sagecal}{{\tt SageCal}}
\newcommand{\sagecalco}{{\tt SageCal-CO}}
\newcommand{\cluster}{{\tt Dawn}}

\newcommand{\nightonetime}{13.0}

\newcommand{\nighttottime}{13.0}
\newcommand{\NCPtotaltime}{800}

\newcommand{\zone}{$9.6-10.6$}
\newcommand{\ztwo}{$8.7-9.6$}
\newcommand{\zthree}{$7.9-8.7$}

\newcommand{\fone}{$121.8 - 134.3$}
\newcommand{\ftwo}{$134.3 - 146.8$}
\newcommand{\fthree}{$146.8 - 159.3$}

\newcommand{\gmcaone}{8/2}
\newcommand{\gmcatwo}{6/2}
\newcommand{\gmcathree}{6/0}

\newcommand{\nightone}{L90490}

\newcommand{\exvarone}{56 \pm 13}

\newcommand{\ulone}{79.6}

\newcommand{\kbest}{0.053}
\newcommand{\ulbest}{\ulone}
\newcommand{\zbest}{\zone}
\newcommand{\exvarbest}{\exvarone}

\newcommand{\ulbestnoise}{57}

\newcommand{\rfiloss}{5}  
 
\newcommand{\rawdatasizeone}{61}
\newcommand{\skymodelsources}{20,800}

\newcommand{\dddir}{122}
\newcommand{\eorFWHM}{10}
\newcommand{\SEFD}{4000}

\newcommand{\polorder}{3}
\newcommand{\fluxlim}{3}

\newcommand{\uvgridsize}{4.58}

\def\ltsima{$\; \buildrel < \over \sim \;$}
\def\lsim{\lower.5ex\hbox{\ltsima}}
\def\gtsima{$\; \buildrel > \over \sim \;$}
\def\gsim{\lower.5ex\hbox{\gtsima}}
\def\ga{\mathrel{\hbox{\rlap{\hbox{\lower4pt\hbox{$\sim$}}}\hbox{$>$}}}}
\def\la{\mathrel{\hbox{\rlap{\hbox{\lower4pt\hbox{$\sim$}}}\hbox{$<$}}}}

\shorttitle{LOFAR 21-cm EoR Power Spectrum Upper Limits}

\shortauthors{Patil et al.}

\begin{document}

\title{Upper limits on the 21-cm Epoch of Reionization power spectrum from one night with LOFAR}
\author{A.H. Patil\altaffilmark{1}, S. Yatawatta\altaffilmark{1,2}, L.V.E. Koopmans\altaffilmark{1}$^{,\dagger}$, A.G. de Bruyn\altaffilmark{2,1},  M. A. Brentjens\altaffilmark{2}, S. Zaroubi\altaffilmark{1,11},  K. M. B. Asad\altaffilmark{1}, M. Hatef\altaffilmark{1}, V. Jeli\'c\altaffilmark{1,8,2}, M. Mevius\altaffilmark{1,2}, A. R. Offringa\altaffilmark{2}, V.N. Pandey\altaffilmark{1},  H. Vedantham\altaffilmark{9,1}, F. B. Abdalla\altaffilmark{7, 13}, W. N. Brouw\altaffilmark{1}, E. Chapman\altaffilmark{7}, B. Ciardi\altaffilmark{4}, B. K. Gehlot\altaffilmark{1}, A. Ghosh\altaffilmark{1}, G. Harker\altaffilmark{3,7,1},
I. T. Iliev\altaffilmark{10},  K. Kakiichi\altaffilmark{4}, S. Majumdar\altaffilmark{12},  M. B. Silva\altaffilmark{1}, G. Mellema\altaffilmark{5},  J. Schaye\altaffilmark{6}, D. Vrbanec\altaffilmark{4},  S. J. Wijnholds\altaffilmark{2}}
\email{$^\dagger$koopmans@astro.rug.nl}

\altaffiltext{1}{Kapteyn Astronomical Institute, University of Groningen, P.O. Box 800, 9700 AV Groningen,The Netherlands}
\altaffiltext{2}{ASTRON, P.O.Box 2, 7990 AA Dwingeloo, The Netherlands}
\altaffiltext{3}{Center for Astrophysics and Space Astronomy, Dept. of Astrophysics and Planetary Sciences, University of Colorado at Boulder, CO 80309, USA} 
\altaffiltext{4}{Max-Planck Institute for Astrophysics, Karl-Schwarzschild-Stra{\ss}e 1, 85748 Garching, Germany}
\altaffiltext{5}{Department of Astronomy and Oskar Klein Centre for Cosmoparticle Physics, AlbaNova, Stockholm University, SE-106 91 Stockholm, Sweden}
\altaffiltext{6}{Leiden Observatory, Leiden University, PO Box 9513, 2300RA Leiden, The Netherlands}
\altaffiltext{7}{Department of Physics and Astronomy, University College London, Gower Street, WC1E 6BT, London, UK}
\altaffiltext{8}{Ru{\dj}er Bo\v{s}kovi\'{c} Institute, Bijeni\v{c}ka cesta 54, 10000 Zagreb, Croatia}
\altaffiltext{9}{Cahill Center for Astronomy and Astrophysics, MC 249-17, California Institute of Technology, Pasadena, CA 91125, USA}
\altaffiltext{10}{Astronomy Centre, Department of Physics \& Astronomy, Pevensey II Building, University of Sussex, Brighton BN1 9QH, UK}
\altaffiltext{11}{Department of Natural Sciences, The Open University of Israel, 1 University Road, PO Box 808, Ra'anana 4353701, Israel}
\altaffiltext{12}{Department of Physics, Blackett Laboratory, Imperial College, London SW7 2AZ, UK}
\altaffiltext{13}{Department of Physics and Electronics, Rhodes University, PO Box 94, Grahamstown, 6140, South Africa}

\begin{abstract}

We present the first limits on the Epoch of Reionization (EoR)  21-cm HI power spectra, in the redshift range  $z=7.9-10.6$, using the Low-Frequency Array (LOFAR) High-Band Antenna (HBA). In total \nighttottime\,h of data were used from observations centred on the North Celestial Pole (NCP). After subtraction of the sky model and the noise bias, we detect a non-zero $\Delta^2_{\rm I} = (\exvarbest {\rm ~mK})^2$ (1-$\sigma$) \exv\ and a best 2-$\sigma$ upper limit of $\Delta^2_{\rm 21} < (\ulbest{\rm ~mK})^2$ at $k=\kbest$\,$h$\,cMpc$^{-1}$ in the range $z=$\,\zbest. The \exv\ decreases when optimizing the smoothness of the direction- and frequency-dependent gain calibration, and with increasing the completeness of the sky model. It is likely caused by  (i) residual side-lobe noise on calibration baselines, (ii) {\sl leverage} due to non-linear effects, (iii) noise and ionosphere-induced gain errors, or a combination thereof. Further analyses of the \exv\ will be discussed in forthcoming publications. 

\end{abstract}



\section{Introduction}\label{sec:intro}

During the Epoch of Reionization (EoR) hydrogen gas in the universe transitioned from neutral to ionized \citep{{1997ApJ...475..429M}}. The EoR is thought to be caused by the formation of the first sources of radiation and hence its study is important for understanding the nature of these first radiating sources, the physical processes that govern them and how they influence the formation of subsequent generations of stars, the interstellar medium (ISM), intergalactic medium (IGM) and black holes; see e.g.\ \citet{Furlanetto:2006bq,Morales:2009p1425, pritchard:2012ja, 2014PTEP.2014fB112N, McQuinn:2015ww} for extensive reviews of the EoR. 

Current observational constraints suggest that reionization took place in the redshift range $6 \lsim z \lsim 10$, with the lower limit inferred from the Gunn-Peterson trough in high-redshift quasar spectra \citep{2001AJ....122.2850B,fan03,fan06}, and the upper limit of the redshift range currently being set by the most recent Planck results,  which yields a surprisingly low value of the  optical depth for Thomson scattering, $\tau_e =0.058 \pm 0.012$ \citep{Planck2016}. This small optical depth  mitigates the tension that exists between the higher optical depth values obtained by the WMAP  satellite \citep{page07, komatsu11, hinshaw13} and the other probes. The current range can easily accommodate photo-ionisation rate measurements \citep{bolton07, calverley11, becker11}, Inter-Galactic Medium (IGM) temperature measurements \citep{theuns02, bolton10, becker13}, observations of high-redshift Lyman break galaxies at $7\lsim z \lsim10$ \citep[see e.g.][]{oesch10, bouwens10, bunker10, bouwens15, robertson15} and  observation of Lyman-$\alpha$ emitters at $z=7$ \citep[see e.g.][]{schenker14, Santos:2016up}. 

It has been long recognized that the redshifted 21-cm emission line provides a very promising probe to observe neutral hydrogen during the EoR \citep[see e.g.][]{madau97, shaver99, furlanetto06a, pritchard12, zaroubi12b}. 

To date, a number of experiments have been seeking to measure this high-redshift 21-cm emission, using LOFAR \citep{haarlem13}, the GMRT \citep{paciga11}, the MWA \citep{Tingay:2013ef, Bowman:2012tx}, PAPER \citep{Parsons:2010kq} and the 21CMA \citep{Zheng:2016vz}. 
These experiments are designed to detect the cosmological 21-cm signal through a number of statistical measures  of its brightness-temperature fluctuations, such as its variance \citep[e.g.][]{patil14, 2014MNRAS.443.3090W} and its power spectrum as a function of  redshift  \citep[e.g][]{morales04, 2005MNRAS.356.1519B,  barkana05,mcquinn06, bowman06,  pritchard07, jelic08, harker09b, harker10, pritchard08}. 

In particular, \citet{jelic08}, \citet{harker10} and more recently \citet{{2013MNRAS.429..165C}, {Chapman:2016jn}} have shown that despite the low signal-to-noise ratio and prominent Galactic and extragalactic foreground emission, the variance and power spectrum of the brightness temperature fluctuations of HI can be extracted from the data collected with LOFAR in about 600 hours of integration time on five fields, barring unknown systematic errors. Deeper integrations on fewer fields can yield similar results\footnote{The power spectrum error scales inverse proportional with the integration time and with the square root of the\textbf{ number of fields}, respectively. This holds in the thermal-noise dominated and low-S/N regime.}. Similar studies have been carried out for the MWA \citep[see e.g.][]{geil08, geil11, 2013MNRAS.429L...5B} and for PAPER \citep[see e.g.][]{Parsons:2012df}. 

\begin{table}
\begin{center}
\smallskip
\begin{tabular}{lll}
\hline
\hline
Phase Centre $(\alpha, \delta; J2000)$     & $0^\mathrm{h}, +90^\circ$ \\
Minimum frequency           & 115.039    & MHz \\
Maximum frequency           & 189.062    & MHz \\
Target bandwidth           &  74.249    & MHz \\
Stations (core/remote)                   & 48 / 13    \\
Raw data volume \nightone & \rawdatasizeone & Tbyte \\
\hline
Sub-band (SB) width          & 195.3125   & kHz \\
Correlator channels per SB        & 64         &     \\
Correlator integration time  & 2          & s   \\
Channels per SB after averaging     &  15, 3, 3, 1         &   \\
Integration time after averaging   & 2, 2, 10, 10         & s   \\
Data size (488 sub-bands)     & 50 & Tbyte \\
\hline
\hline
\end{tabular}
\end{center}
\caption{Observational and correlator set up of LOFAR-HBA observations of the North Celestial Pole (NCP).}
\end{table}\label{tab:dataproc}

At present, a number of upper limits on the brightness-temperature power spectrum  have been published. \citet{paciga13} have used the GMRT to set a 2-$\sigma$ upper limit on the brightness temperature at $z=8.6$ of $\Delta^2_{21} < (248\, \mathrm{mK})^2$  at wave number $k\approx 0.5$\,\cmpc. \citet{Beardsley:2016tp} provided a 2-$\sigma$ limit at $z=7.1$ of $\Delta^2_{21} < (164\, \mathrm{mK})^2$ at $k \approx 0.27$\,\cmpc\ from MWA. The PAPER project provided the tightest upper limit yet of $\Delta^2_{21} < (22.4\, \mathrm{mK})^2$ in the wave number range $0.15\le k \le 0.5$\,\cmpc\ at $z = 8.4$ \citep{ali15}.

\begin{figure*}\label{fig:FieldImage}
\centering
\hspace{1cm}
\includegraphics[width=0.8\hsize]{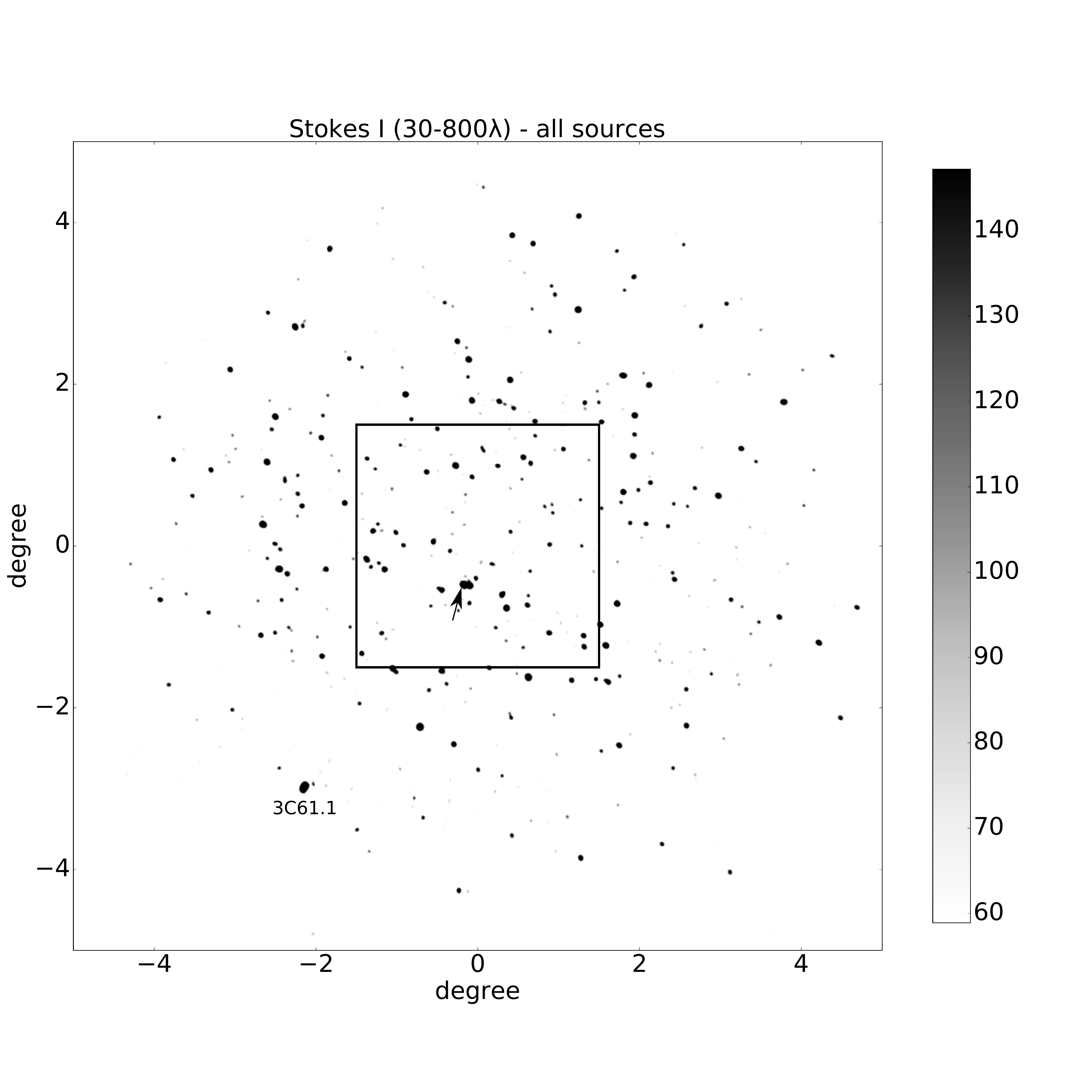}
\vspace{-1.0cm}
\caption{A relatively narrow-band continuum (134.5-137.5\,MHz) LOFAR-HBA image of $10^\circ \times 10^\circ$ of the North Celestial Pole (NCP) field, centered at dec +90.0$^\circ$. Baselines between 30-800$\lambda$ were included, using uniform weighting. No sources have been subtracted and the image is cleaned to a level sufficient to show the brightest few hundred sources above 60~mJy. The $3^\circ\times3^\circ$ box delineates the area where we measure the power spectra. The bright extended source in the lower-left is 3C61.1 (J0222+8619), discussed in the text. The bright (7.2\,Jy) compact source near the NCP is indicated by an arrow. The intensity units are mJy/PSF (see text). Right Ascension increases clockwise; RA=00h is towards the bottom.}
\end{figure*}

Here we report the first 21-cm EoR power-spectrum limits from the LOFAR EoR Key Science Project (KSP) based on a single night of data set acquired in the first LOFAR observing cycle (i.e.\ Cycle-0).  The approach taken in the LOFAR EoR project differs in two important aspects from those in the other experiments mentioned above. Firstly, in order to remove the chromatic response from the multitude of bright continuum sources found in a typical LOFAR observation we have developed a comprehensive sky model. This model is then used to calibrate the data in a large number of directions. We then also remove these sources and their responses from the visibility data. Secondly, we use a technique that goes by the name of Generalized Morphological Component Analysis (GMCA, hereafter {\tt GMCA}) to remove the residual compact and remaining diffuse foregrounds. Both aspects, as applied to real data, have not been described in detail before. We will therefore describe these processing steps, and how we have arrived at the chosen parameters and strategy, in some detail. 

This paper is organized as follows. In Sect.~\ref{sec:observations} we describe the observational set-up and the data that is being analysed. In Sect.~\ref{sec:dataproc} we describe the various steps in our data-processing.   In Sect.~\ref{sec:calibration} we describe the calibration of our data. In Sect.~\ref{sec:imaging} our imaging procedures are described. The resulting  power spectra are presented in Sect.~\ref{sec:results}. The paper concludes with a summary and outlook in Sect.~\ref{sec:conclusions}. We assume the standard cosmology \citep{2015arXiv150201589P} and scale the Hubble constant as $h=H_0/$100~km\,s$^{-1}$\,Mpc$^{-1}$.
  

\section{Observations}\label{sec:observations}

The observations conducted for the LOFAR EoR project are concentrated on two windows: the North Celestial Pole (NCP) and the bright compact radio source 3C196 \citep[see][]{2010A&A...522A..67B, 2013A&A...550A.136Y}.  The results presented in this paper are based on data taken on the NCP field with the LOFAR telescope \citep[][] {2013A&A...556A...2V} in the night from 2013 February 11/12. The frequency range from 115 to 189 MHz was covered using receivers in the so-called LOFAR-HBA band (where HBA refers to High Band Antenna).  All 61 Dutch LOFAR-HBA stations \citep[e.g.][and Table~\ref{tab:dataproc}]{2013A&A...556A...2V} available in early 2013, participated in the observations.

\subsection{Data Sets}

NCP observations are usually scheduled from ``Dusk to Dawn'', and have typical durations of 12--15.5\,h during the Northern hemisphere winter.  The phase and pointing centre was set at RA=0$^{\rm h}$, DEC=+90$^\circ$ (Table\,\ref{tab:dataproc}).  The NCP can be observed every night of the year making it an excellent EoR window.  Currently $\sim$\NCPtotaltime\,h of good-quality data have been acquired during Cycles 0--5\footnote{\url{http://www.astron.nl/radio-observatory/cycles-allocations-and-observing-schedules/cycles-allocations-and-observing-schedu}}, under generally good ionospheric conditions \citep[see e.g.][]{Mevius:2016ba} and in a moderate RFI environment \citep[e.g.][]{2012arXiv1210.0393O}. We refer to \cite{2013A&A...550A.136Y} for a detailed description of the NCP field and early LOFAR commissioning observations. 

For the analyses presented in this paper, a single \nightonetime-hr data set (i.e.\ \nightone) was selected from a larger set ($\sim$150\,h of data) that was previously analysed with an earlier version of the calibration code \sagecal\ \citep{2011MNRAS.414.1656K}. The data in this night is of excellent quality, based on the Stokes V rms noise, RFI levels and ionospheric conditions. We recently re-processed this dataset using an improved calibration strategy \sagecalco\ \cite[see Sect.~\ref{sec:dataproc};][]{Yatawatta:2015gs, Yatawatta:2016we} yielding a more robust calibration than previously \cite[used in e.g.][]{2013A&A...550A.136Y}.

\subsection{Station Hardware and Correlation}

The LOFAR array has a rather complex, hierarchical configuration. Here we give a brief summary, restricting ourselves to the HBA-band configuration in which we recorded our data.  For a more detailed description of LOFAR hardware we refer to \citet[][] {{2013A&A...556A...2V}}.

Individual HBA-dipoles are grouped in units of 4$\times$4 dual-polarisation dipoles. This unit is called a tile. It has a physical dimension of 5$\times$5~m. The 16 dipole signals are combined in a summator, an analogue beam-former, the coefficients of which are regularly updated when we track a source. In the case of the NCP this is not needed. A core station (CS) consists of 24 closely packed tiles; a remote station (RS) has 48 tiles. The core stations are distributed over an area of about 2~km diameter, in co-located pairs of stations which share a receiver cabinet. The remote stations are spread over an area of about 40~km East-West and 70~km North-South.  Although all remote stations have 48 tiles we only used the inner 24 tiles in the beam-former in order to give both core and remote stations the same primary beam. The receivers at a LOFAR station digitize the data at 200\,MHz clock speed, fully covering the frequency range from 100--200\,MHz \citep[][] {{2013A&A...556A...2V}}. This produces 512 sub-bands of each 195\,kHz bandwidth. The fibre network used to bring signals from the stations to the correlator can transport a maximum of 488 of these 512 sub-bands. The correlator is located at the computing centre at the University of Groningen, about 40~km north of the LOFAR core. We therefore record a total RF bandwidth of 96\,MHz \citep[][]{{2013A&A...556A...2V}}. Of this bandwidth 74\,MHz, i.e. all frequencies between 115 and 189\,MHz, was allocated to the target field. The remaining 22\,MHz were distributed, sparsely covering the same frequency range, over a hexagonal ring of six flanking fields located at an angular distance of 3.75$^\circ$ from the NCP.  The  flanking-field data are used for calibration purposes, ionospheric studies and construction of models for sources located at the edges of the station (primary) beam.  In the LOFAR EoR observations the correlator generates 64 frequency channels, each of 3.1\,kHz, per sub-band and stores the visibility data at 2\,s time resolution in so-called Measurement Sets (MS). Every sub-band is stored in a separate MS. 

\subsection{Intensity scale and Noise}

The intensity scale in the data is set by the flux density of the very compact  source located at RA=01h17m32s, Dec=89$^\circ$28$'$49$''$ (J2000). From (unpublished) European-scale LOFAR long baseline data this source is found to have a size of about 0.3$''$ and is therefore completely unresolved on the Dutch LOFAR baselines used in this work. Following calibration against 3C295 \citep{2012MNRAS.423L..30S} we find the source to show a spectrally broad peak at 7.2\,Jy in the range from 120--160~MHz. Note that this is its apparent flux at 31 arcmin from the pointing centre which is at DEC=90$^\circ$. However, the source bends down at frequencies below 100\,MHz and above 200\,MHz.  We have adopted a constant flux density over the frequency range for which we show data in  this paper. We estimate this value to be good to 5\% on the flux scale of \citet{2012MNRAS.423L..30S}. This flux density is about 30\% larger than adopted in \cite{{2013A&A...550A.136Y}}, where we presented the first NCP observations with LOFAR-HBA. 

The thermal noise in the data is determined using the temporal statistics of the real and imaginary parts of the XY and YX visibilities in narrow 12\,kHz channels. These are observed to be Gaussian distributed. The narrow-band visibility noise also correctly predicts the narrow-band image noise as determined from differences between naturally weighted images in all Stokes parameters. At this spectral resolution broad-band instrumental and ionospheric errors indeed cancel almost perfectly. The measured visibility noise implies a System Equivalent Flux Density (SEFD) of $\sim$\SEFD\,Jy per station, which is close to the  expected value in the direction of the NCP, after correcting for the beam gain away from the zenith \citep[see][for the zenith SEFD values]{2013A&A...556A...2V}. 

We note that when we quote peak flux densities of sources, or noise levels in images, we will give them as flux density per synthesized resolution element. This is what is normally called the Point Spread Function (PSF). This convention therefore differs from the terminology used in radio astronomy, which is to quote fluxes per beam. However, phased arrays, such as LOFAR, have a time-variable (primary) beam which has often lead to confusion. So to be precise, when we refer to flux density per PSF, we refer to the flux density per solid angle as subtended by the PSF. For a Gaussian PSF, as is often used in restored images, the relevant solid angle would then be equivalent to 1.13 times the square of the Full Width at Half Maximum (FWHM) of the PSF.


\section{Data Processing}\label{sec:dataproc}
\subsection{Compute and Storage Resources}

Processing a single 13-hr LOFAR-HBA data set is computationally expensive and currently takes $\sim$50\,h on a dedicated compute-cluster consisting of 124 {\sl NVIDIA} K40 GPUs, called \cluster\footnote{\url{https://www.astron.nl/sites/astron.nl/file/cms/201500151-01_Astron_News_Winter_2015-01_\%5BWeb\%5D.pdf}} hereafter. Most of the processing time is needed for the calibration, specifically the direction-dependent calibration (see Sect.~\ref{sec:calibration}). The imaging step is computationally negligible. We are working on further optimization and automation of the calibration. All data processing on the visibilities is done on \cluster, located at the Centre for Information Technology\footnote{\url{http://www.rug.nl/society-business/centre-for-information-technology/}} of the University of Groningen. Petabyte-storage is distributed over \cluster, a dedicated storage cluster at ASTRON\footnote{\url{ http://www.astron.nl/}} and at various locations of the LOFAR Long-Term-Archive (LTA).

The LOFAR-EoR data processing pipeline -- prior to power-spectrum extraction (Sect.~\ref{sec:results}) -- consists of a large number of steps: (1) Preprocessing and RFI excision, (2) data-averaging, (3) direction-independent calibration (henceforth \dical), (4) direction-dependent calibration (henceforth \ddcal) including sky-model subtraction, (5) short-baseline imaging, (6) removal of residual foregrounds. In this section we describe the hardware and software used in steps (1) and (2). The calibration of our data, steps (3) and (4), are described in detail in Sec~\ref{sec:calibration}.  All data-processing codes are publicly available and links to the source codes and documentation are given where applicable.

\begin{table}
\begin{center}
\smallskip
\begin{tabular}{lll}
\hline
\hline
{\bf Parameter }    & {\bf Value}  & {\bf Comments} \\
\hline
Sky-model components             & $\sim$\skymodelsources   & Compact \\
Flux-limit sky model             & $\sim$\fluxlim ~mJy  \\
Order $P^S_n$ source spectra     & \polorder & Polynomial \\ 
\hline
DI-Calibration directions         &  2 \\
DD-Calibration directions         & \dddir   & Source \\
& & clusters \\
Calibration baselines             &  $\geq$250~$\lambda$ \\
Order $B^G_n$ gain regul.         & 3 & Bernstein \\
& & Polynomial \\ 
Solution interval                 & 10 min \\
\hline
$uv$-grid cells                 & $\uvgridsize \times \uvgridsize$~$\lambda$\\
$w$-slices                      & 128 \\
\hline
EoR Imaging baselines          & 50-250~$\lambda$  \\
EoR Imaging FoV                & $3^\circ \times 3^\circ$  \\
EoR pixel size                 & $0.5'\times 0.5'$ \\
EoR Imaging Resolution         & $\sim$\eorFWHM' &  FWHM \\
EoR Freq. Resolution           & $\sim$60~kHz \\
\hline
\hline
Redshift range \#1 & \zthree  \\
Freq. range & \fthree~MHz  \\
{\tt GMCA} components & \gmcathree & Stokes I/V. \\
\hline
Redshift range \#2 & \ztwo & \\
Freq. range & \ftwo~MHz  \\
{\tt GMCA} components & \gmcatwo & Stokes I/V. \\
\hline
Redshift range \#3 & \zone &  \\
Freq. range & \fone~MHz \\
{\tt GMCA} components & \gmcaone & Stokes I/V. \\
\hline
\hline
\end{tabular}
\end{center}
\caption{Calibration and sky-model parameters and settings.}
\end{table}\label{tab:calsetup}

\begin{figure*}\label{fig:panelIVsub}
\centering
\includegraphics[width=0.8\hsize]{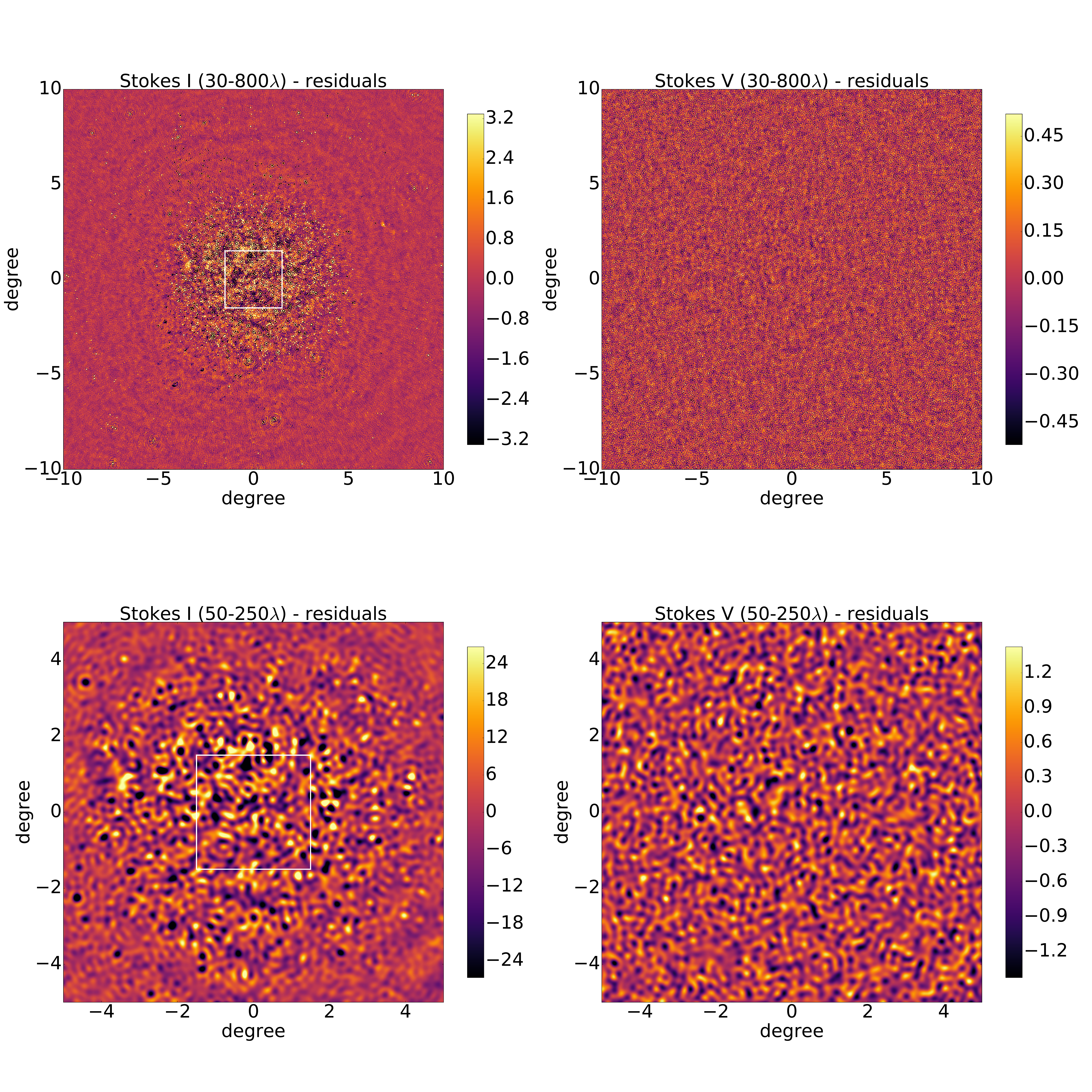}
\caption{Stokes I and Stokes V images after sky-model subtraction for the baseline ranges 30-800$\lambda$ (top panels) and 50--250$\lambda$ (bottom panels). Sub-bands with frequencies between 121 and 134 MHz went into these images. Note the reduction in the displayed field-of-view from 20$\times$20$^\circ$ to 10$\times$10$^\circ$. Intensity units are in mJy/PSF and the scale range is set by plus and minus three times the standard deviation over the full field in all images. Note the  noise-like structure in the two Stokes V images. i.e. a lack of any features. The Stokes I images, on the other hand, clearly show the LOFAR-HBA primary beam attenuation effects on the remaining diffuse emission. The level of this emission is limited by the classical confusion noise within the primary beam. The $3^\circ \times 3^\circ$ box delineates the area where we measure the power spectra.}
\end{figure*}

\subsection{Pre-processing, RFI excision and Data averaging}

Standard (tabulated) corrections are applied to the raw visibilities (e.g.\ flagging of known bad stations or baselines) using {\tt NDPPP}\footnote{\url{http://www.lofar.org/operations/doku.php?id=public:user_software:ndppp}}.
RFI-flagging is done on the highest-resolution data using the {\tt AOflagger}\footnote{\url{https://sourceforge.net/p/aoflagger/wiki/Home/}} \citep{2012A&A...539A..95O} and leads to a typical loss of $\sim$\rfiloss\% of the LOFAR-HBA $uv$-data. 

Several clean data products at different temporal and frequency resolutions are then created. We first flag channels 0, 1, 62, and 63 at the edges of the sub-bands to avoid low-level aliasing effects from the poly-phase filter used to provide the fine frequency resolution. The remaining 60 channels are averaged to 15 new channels each of 12\,kHz. These data are archived for later analysis (to search for 21-cm absorption in bright sources and permit searches for fast transients). We then further average the data to 3 channels each of 61\,kHz  while maintaining the 2\,s time resolution. At this resolution the time and frequency smearing of off-axis sources is still acceptable at the longest baselines.  This is important for high-resolution source modelling (see Sect.~\ref{sec:skymodel}). For initial calibration, we also formed a low-resolution product with a temporal resolution of 10\,s (see Table~\ref{tab:dataproc}). We note that in our previous analysis of the NCP field \citep{{2013A&A...550A.136Y}} we used a spectral resolution of 183\,kHz, i.e. a full sub-band, in the processing. Currently we conservatively flag baselines between stations that share a common electronics cabinet, to avoid any correlated spurious signals.  There are 24 such baselines in the LOFAR core. These station pairs have projected baselines between about 40 and 60$\lambda$, depending on frequency. We expect to recover most of these data in forthcoming analyses, potentially increasing the number of short baselines by up to a factor $\sim$2.5 in that range.

\subsection{The NCP Sky Model}\label{sec:skymodel}

The continuum foreground for EoR-experiments consist of two distinct components \citep{shaver99}. On very short baselines, less than about 10\,$\lambda$, the diffuse Galactic synchrotron emission  starts to dominate the visibilities. Also the intense emission of Cas-A and Cyg-A, the two brightest radio sources in the Northern hemisphere located in or close to the Galactic plane, and very far from our EoR windows, occasionally enters a distant side-lobe and will then dominate the visibilities.  The shortest baseline in LOFAR is about 35~m and corresponds to about 15--20$\,\lambda$. This means that the diffuse Galactic  component is a) hardly detectable in our data, and, b) also very difficult to model.  The more problematic component, and the one dominating our images are the extragalactic sources. Most of these have an angular size less than a few arcminutes.  Source model components are determined from the highest resolution LOFAR images which have an angular resolution of $\sim 6$\,arcsec FWHM.  For some of the brightest sources we have also made use of international baselines in LOFAR,  which provide a resolution down to 0.25\,arcsec. 
The discrete source model for the NCP field has been iteratively built up over the last several years, using a program called {\tt buildsky}\footnote{Included in the {\tt SageCal-CO} repository:\\ \url{https://sourceforge.net/projects/sagecal/}} \citep[see e.g.][]{2013A&A...550A.136Y}. Fig.~\ref{fig:FieldImage} shows a 3\,arcmin resolution 10$^\circ$$\times$10$^\circ$ image of the NCP. It reveals  the brightest few hundred sources  down to a flux density limit of 60~mJy.  Our sky model includes sources up to 19\,deg distance from the NCP, excluding Cas-A and Cyg-A which are much further away. In fact all sources that are bright enough to cause (chromatic) side-lobes in  the  inner few degrees of the field were included in our model. We expect this  model will continue to grow in the next year when we expect to go deeper. The current calibration sky model (Stokes I) consists of $\sim$\skymodelsources\ unpolarized source components, including Cas-A and Cyg-A (see Table\,\ref{tab:calsetup}).  It has components down  to $\sim$\fluxlim\,mJy, i.e. the apparent flux in our model, which are modelled either as a point-source, multiple Gaussians or shapelets. Each source has a smooth frequency model (polynomial of order \polorder) which is regularly updated, as data is combined and calibration improves. Although sources down to a few mJy were included in our sky model, our low resolution residual images (see Sect.~\ref{sec:imaging}) still show many positive and negative sources with fluxes going up to +50 and -50~mJy. These are located near the brighter sources in the field which still leave residuals following the calibration.    

\section{Calibration}\label{sec:calibration}
 
Our calibration strategy has been developed over a period of several years. In this period we have explored a wide set of processing parameters the choice of which was guided by a combination of information-theoretical arguments, end-to-end simulations, a thorough analysis of the image cubes and the effects of unmodelled structure.  To give some insight into the problem we will start with an outline of our calibration strategy. 

The NCP field is dominated by two bright sources (see Fig.~\ref{fig:FieldImage}. One of them (J0117+8928) is compact and has a flat spectrum (see Sect~\ref{sec:skymodel}) and is located only 31$'$ from the pointing and phase centre of the observation. The other source (3C61.1; J0222+8619, an FR-II radio galaxy) is located at the edge of the primary field of view. It has a complex morphology with both intense sub-arcsecond as well as arcminute scale structure. 
However, the most problematic aspect of 3C61.1 is its location close to the first null of the primary beam for the highest frequencies used in this analysis. Because the LOFAR-HBA core station primary beam is much larger at 115\,MHz than at 177\,MHz, 3C61.1 dominates the visibilities at frequencies below 130 MHz. In fact, the source reaches an apparent flux density of $\sim$ 14~Jy at 115~MHz.  The ionospheric phase delays will therefore be dominated by those present towards 3C61.1.  This frequency-dependent behaviour is exacerbated by the imperfect knowledge of the beam-gains of the 61 stations close to the edge of the primary beam. The combination of the properties of 3C61.1 forced us to depart from the normal two-step calibration of LOFAR data, which consists of a \dical, followed by a \ddcal. In essence, our \dical\ is now done towards two directions simultaneously. We use \sagecalco\ for both calibration steps. This is a relatively recent departure of the calibration procedure adopted in the past. The main reason is to make the direction-independent calibration solutions independent of those found towards the bright problematic source 3C61.1. However, to not unnecessarily complicate the description below we will continue to refer to this first step as \dical.  Table~\ref{tab:calsetup} lists the most relevant calibration parameter settings.

\subsection {Direction-independent calibration}

The \dical\ is done at 61\,kHz frequency resolution and  2\,s time resolution using all baselines in the array. The sky models for the two directions consist of (i) all sources in the field, dominated by the compact 7.2~Jy source near the centre, except 3C61.1, and (ii) the source 3C61.1 itself.  In this first step the fast ionospheric phase variations towards the two brightest sources can be solved for.  The S/N ratio per sub-band is sufficiently high to work at this high time resolution. We solve for the gains per sub-band of 183\,kHz, but use the full frequency domain \citep[see for details][]{Yatawatta:2015gs, Yatawatta:2016we} to fit for the slow as well as fast variations in frequency. This \dical\ will absorb the structure in the band-pass response of the stations. This structure is due to low-pass and high-pass filters in the signal chain as well as reflections in the coax-cables between tiles and receivers \citep[see e.g.][]{{2012arXiv1210.0393O}}. In the LOFAR core stations the antennae and receivers are connected via 85~m coax-cables. These cause a 920~ns delayed signal with a relative intensity of -22dB. This causes a $\approx~1\%$ ripple  in the gains  with a  a periodicity of 1.09 MHz.  These frequency-ripples are similar for all core stations. The remote stations, on the other hand, have features at 1.09~MHz and 1.38~MHz because two sets of coax-cables with lengths of 85~m and 115~m are used. The frequency-dependent station gains and ionospheric delays found towards 3C61.1 in this first calibration step {\sl therefore do not influence the gain solutions for the other direction}.  Finally, we correct the visibilities for the gains found for the full field. Note that we do not yet remove 3C61.1 from the data in this \dical\ step. 

\subsection{Direction-dependent calibration}

We want to create a field of view -- from which we want to extract the power spectra -- free from as many sources and their artefacts as possible. Most of the bright sources are distributed over an area of about 8$^\circ$ diameter (see Fig.~\ref{fig:FieldImage}) but sources with apparent flux densities down to 3~mJy are found out to radii of at least 10$^\circ$. Over such a large area the station-beam gains vary enormously and unpredictably (in detail). Also the ionospheric iso-planatic angle is expected, and indeed observed, to be typically 1--2$^\circ$. To remove all these sources will therefore require \ddcal. Hence \ddcal\ is always associated with subtraction of the sky model. We do not replace these sources in our image cubes with their model (as is often done in {\tt Cleaning}). We had to find a compromise between the number of directions to solve for beam and ionospheric errors, the maximum baseline to use in calibration, the time scale on which to solve for station gains and ionospheric phases, on the one hand, and the number of constraints provided by the data, on the other. Long baselines  provide the most constraints. However, by using long baselines, up to a projected maximum baseline of 70~km, we are vulnerable to ionospheric and sky-model errors. Whereas \ddcal\ is obviously important, the very large number of parameters for which we have to solve also can lead to ill-conditioning of the problem. This has led to a range of subtle and less subtle consequences, which we will describe below. 

\ddcal\ is an iterative process described in more detail in \cite{{2015MNRAS.449.4506Y}, Yatawatta:2016we}. We group the sky-model components in \dddir\ directions, called source ``clusters" \citep[][]{{2013MNRAS.430.1457K}}. Most clusters will have a large number of components although its response might occasionally be dominated by a single source. Clusters are typically 1--2 degrees in diameter. \sagecalco\ uses an expectation maximization (EM) algorithm to solve for the four complex gains (full Stokes) in one effective Jones matrix per direction \citep[see e.g.][]{{Hamaker:1996p1468},{2011A&A...527A.106S}}. This Jones matrix describes the combination of all direction-dependent effects (i.e.\ beam errors, ionospheric phase fluctuations, etc.) and is assumed to be the same for all sources in a cluster. We plan to relax this assumption in the future.    

\begin{figure*}\label{fig:freqslice}
\centering
\includegraphics[width=0.9\hsize]{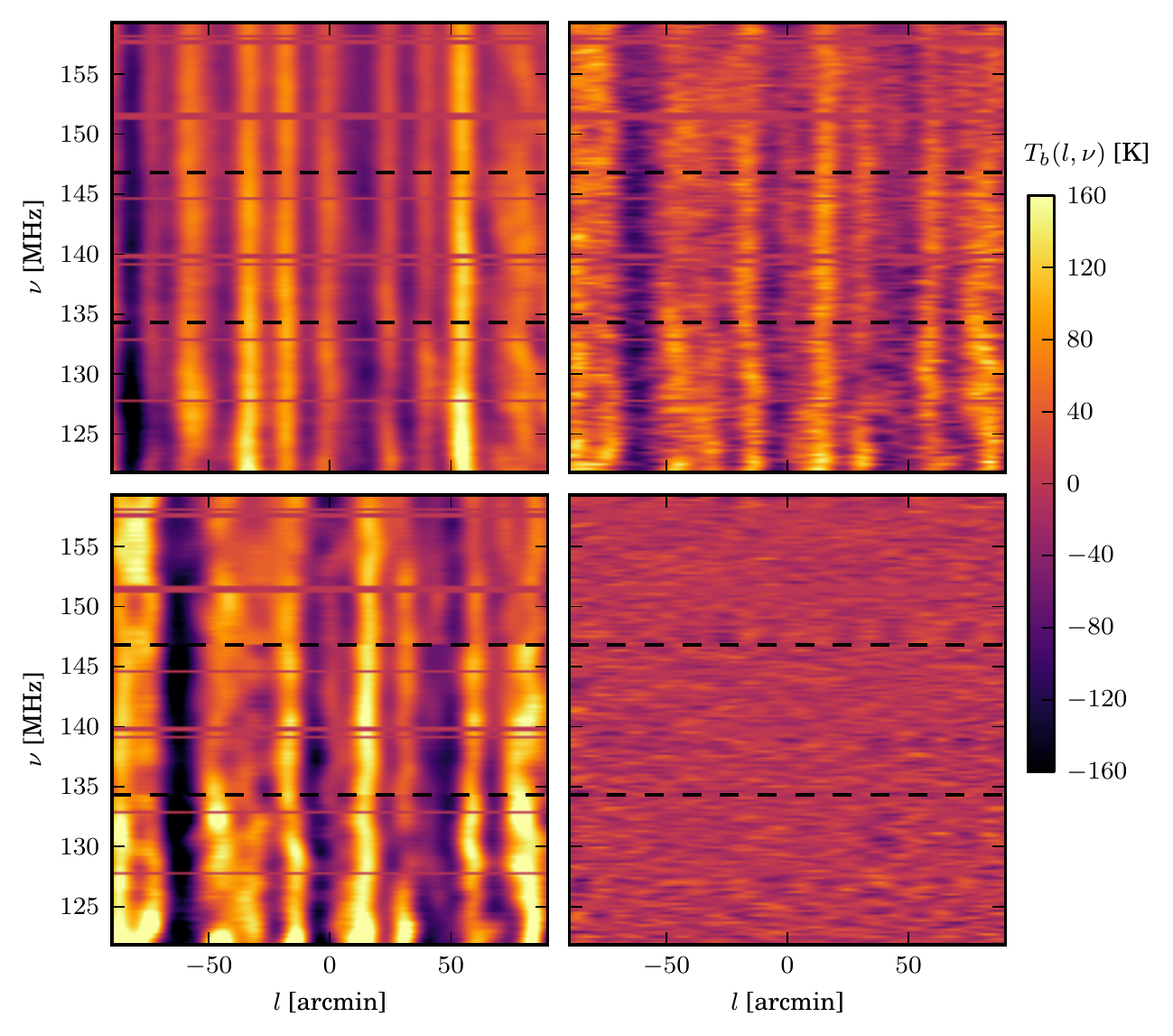}
\caption{{A slice across the centre of the 50--250$\lambda$ Stokes-I data cube along the frequency direction. Top left: slice after \dical\ with only 3C61.1 subtracted;  the intensity scale, converted to brightness temperature, refers to this panel.  Top right: after \ddcal\ where the calibration sky model, consisting of compact sources, is subtracted with their respective direction-dependent gain solutions. The intensity scale is now multiplied by 10 for improved visualization; Bottom left: {\tt GMCA} model (scale also multiplied by 10); Bottom right: {\tt GMCA} residuals (scale multiplied by another factor of 20). The red horizontal bands are due to data lost due to RFI-flagging. The black dashed lines border the three redshift ranges. Note the factor $\sim$200 reduction in intensity after {\tt GMCA}.}}
\end{figure*}

The complex gains are solved for all clusters simultaneously. We use a third-order Bernstein polynomial basis function \citep{{Yatawatta:2016we}} in the frequency direction as a regularisation {\sl prior} on the gain solutions over the full bandwidth.
Hence, although the gains are allowed to deviate from the smooth prior, this will be penalized by a quadratic regularization term  \citep[i.e.\ penalty function; see][]{Yatawatta:2016we}. The regularization constant is optimized to minimize the mean squared error between the gain solutions per sub-band and the smooth third-order Bernstein polynomial basis function. If the regularisation constant is chosen too large, the data cannot be fitted, and if chosen too small, the data are over-fitted. This fitting process is iterated typically $\sim$30 times, simultaneously optimising the weights of the Bernstein polynomial basis functions and the individual gains for all \dddir\ directions and for all sub-bands (i.e.\ 195\,kHz). The solutions are applied to the separate narrow 61\,kHz channels), until convergence is reached. 

The solution time intervals are dependent on the strength of the signals in the various clusters and vary between 1 and 20~minutes. This time-scale should be sufficient to fit for the slowly varying station-beam gain variations. However, 20 minutes is too long to capture ionospheric phase variations on most baselines.  The isoplanatic angle in a typical LOFAR observation in the HBA-band is typically 1--2$^\circ$. Many of the relatively bright radio sources in the field, and especially those that are not dominating the cluster they are assigned to, will then be imperfectly calibrated and leave residuals.  An imperfect calibration of these sources, however, will also influence the gains for the stations involved in the short baselines on which we are most sensitive to EoR signals. This could lead to baseline-dependent decorrelation effects. How these effects manifest themselves in the final residual data on the shortest baselines is still under investigation \citep[see e.g.][]{2016MNRAS.tmp..226V}. We expect to reduce the \sagecalco\ solution time in the future and also use separate solution intervals for amplitude and phase.

\subsection{Suppression of Diffuse Emission}\label{sec:suppression}

{\sl DD}-calibration can remove diffuse structures (i.e.\ power) in Stokes I, Q and U. This has been discussed and documented in detail  in \cite{Patil:2016td}. Because our calibration sky model only consists of relatively compact sources, this removal of diffuse emission occurs because of a ``conspiracy'' of the direction-dependent gains -- or equivalently the direction-dependent PSFs -- convolving the sky model with extended low-level PSFs and removing structures in the data that are not part of the sky model. Whereas using too few calibration directions leaves artefacts around compact sources, using too many will remove structure \citep{Patil:2016td}. This is opposite (not in contradiction) to the issue noted by \citet{2016MNRAS.461.3135B}, where an incomplete/inaccurate sky model in MWA data simulations causes gain errors on all baselines which then leads to \exv\ in the EoR 21-cm power spectrum. 
To mitigate both problems we split the baseline set into non-overlapping calibration and EoR-imaging sub-sets, with a cut at several hundred $\lambda$, beyond which we see no evidence for diffuse emission in Stokes I, Q and U. We calibrate using the longer baselines and we analyse the EoR signal on the shorter baselines.  Furthermore we use our high-resolution images to create a sky-model that reaches well below the classical confusion noise level corresponding to the resolution of the $50-250\lambda$ baselines (see Sect.~\ref{sec:skymodel}; Fig.~\ref{fig:panelIVsub}).  We have tested the effects of both higher and lower cuts. The chosen cut of 250$\lambda$ is the compromise adopted in our current processing.  This value remains well above the baseline lengths where (realistically speaking) LOFAR could detect an EoR signal. 

We note that {\sl if} diffuse emission can be included in the model, the baseline cut may not be needed. This is still under investigation and some encouraging results have already been obtained.

\subsection{Excess Noise}\label{sec:exn}

Whereas an imposed baseline-cut largely resolves the issue of suppression of diffuse emission, it leads to  \exn\ on the short (imaging) baselines \citep[see][their Figures~11 and 12]{Patil:2016td}, while simultaneously decreasing the noise and unmodelled flux on long (calibration) baselines. This discontinuous change in the noise level, at the location of the uv-cut, is absent when we calibrate using {\sl all} baselines as we did in our original calibration strategy. 
Extensive simulations show that this \exv\ on the short baselines that are excluded in the calibration can be caused by three effects \citep[e.g.][]{Patil:2016td}: 

\paragraph{Leverage}

Leverage is an effect known in signal processing when a data set is calibrated using only a subset of the data. This leads to an increase of variance on the excluded baselines and a decrease on those that are included \citep[see appendix in][for a mathematical description]{Patil:2016td} and is related to a bias introduced in non-linear optimization \citep{Laurent:1992ci, COOK:1986cg}.

\paragraph{An Incomplete or Inaccurate Sky Model}

Even on the long baselines, where we are not limited by classical confusion, the sky model remains incomplete and imperfect. This is partly due to our inability to determine accurate source parameters for sources with an angular size equal to the PSF. Another important source of errors in source models is related to differential ionospheric corruptions across the source clusters used in \sagecal. The spectrally complex model of the brightest source (at frequencies below 130\,MHz) in the field, 3C61.1 (Fig.\ref{fig:FieldImage}), still needs improvement using sub-arcsecond structural information from the European $\sim$1000~km baselines now available.  The chromatic residual side-lobe noise from all these imperfectly calibrated sources, will affect the frequency-dependent gain solution on a frequency scale that depends on the distance of the source from the phase centre \citep[see e.g.][]{Patil:2016td, 2016MNRAS.461.3135B}.

\paragraph{Signal-to-Noise} 

Using fewer and only longer baselines increases the thermal and ionospheric speckle noise \citep{{2016MNRAS.tmp..226V}}, and hence the resulting gain errors. We think this effect is still the smallest of the three although it can interact or be amplified by the first two effects, especially when the optimization problem is ill-conditioned. We note, however, that \sagecalco\\ includes regularisation to suppress the latter \citep[see][for a detailed analysis]{Yatawatta:2016we}.\\

\subsection{Regularisation of complex direction-dependent gains}

The three effects described in Sect.~\ref{sec:exn} lead to additional spectral fluctuations on short baselines \citep[see][]{Patil:2016td}. 

To mitigate the amplification or propagation of small (non-instrumental) gain fluctuations, we penalize irregular gain solutions via a regularisation function \citep[][]{2015MNRAS.449.4506Y, Yatawatta:2016we}. We use a Bernstein polynomial of third order as prior on the {\sl DD}-gain solutions \citep[see e.g.][]{Farouki:2012iz}. 
\ddcal\ over the full frequency domain, splitting the calibration and imaging baselines, using a detailed sky model and regularizing the gain solutions, are all currently combined in the single framework of {\tt SageCal-CO}\footnote{\url{https://sourceforge.net/projects/sagecal/}} and runs efficiently on the parallel cluster \cluster, using {\tt MPI} and {\tt CUDA}.

\subsection{Sky-model Subtraction and Gridding}

Rather than correcting the $uv$-data (or images) for direction-dependent gain errors, we subtract the sky model from the visibility data in \sagecalco\ using their full-Stokes gain solutions. We use the regularised gain solutions per sub-band/channel rather than the Bernstein polynomial itself, which is purely used as a prior function for the gains (in the case of very strong regularisation these two gain solutions as function of frequency would become identical).  Subtraction of the sky model also removes their polarization leakage from Stokes I to Stokes Q, U and V \citep[see e.g.][]{{Asad:2015vv}, 2016arXiv160404534A}, as well as their beam and ionospheric effects, but only on spatial scales of the cluster diameters and their respective solution time-intervals, or larger. 
Subsequently the $uv$-data inside the 50--250$\lambda$ annulus is gridded using $\uvgridsize \lambda \times \uvgridsize\lambda$ $uv$-cells and 128 $w$-slices, using a prolate spheroidal wave-function kernel \cite[see][for details]{2010arXiv1008.1892Y, Noorishad:2011eg}. 
%


\section{Image Cubes}\label{sec:imaging}

We make use of a GPU-enabled imager called {\tt ExCon}\footnote{\url{https://sourceforge.net/projects/exconimager/}}, which can optimize the visibility weights to minimize the spectral dependency of the PSF \citep{Yatawatta:2014hn}. A spectrally-independent PSF improves the performance of {\tt GMCA}. We also have used WSClean \citep{offringa-wsclean-2014}, and its image deconvolution features, for general verification of our images.

\subsection{Residual-Image Cubes}

We produce $3^\circ\times3^\circ$ image cubes with $0.5'\times0.5'$ pixels using the 50-250$\lambda$ baselines, for the frequency ranges \fone\,MHz, \ftwo\,MHz\ and \fthree\,MHz, respectively (see Table~\ref{tab:calsetup}). We do not apply a correction for the slowly varying station beam in the imager. These image have a PSF of $\sim$\eorFWHM\ arcmin FWHM. The spectral resolution of the cubes, in all four Stokes parameters, is 61\,kHz. We use Stokes V as a measure of the data-quality and noise level. Note that \ddcal\ only removes the discrete  source components in each source cluster using the complex gain corrections derived for that direction. That is, the residual images for all cubes processed from this point onwards have only \dical\ applied to them.   Table~\ref{tab:calsetup} lists the most relevant imaging parameter settings.  

In single-night integrations we have found evidence for very faint non-celestial signals in only a dozen subbands, concentrating near the NCP. Such signals could be caused by faint stationary RFI or low-level but stable cross-talk in the system. Any stationary (w.r.t.\ the array) RFI sources would coherently add at the NCP (i.e.\ their side-lobes rotate as the sky rotates and add coherently only on the NCP). The absence of such RFI signatures is a good sign of high data fidelity. Note that strong RFI was already flagged using {\tt AOFlagger}; \cite{{2012A&A...539A..95O}}.  
\nightone\ is ionospherically well-behaved with diffractive scales of 21, 12, 18 km, respectively, in consecutive $\sim$4-hr time ranges \citep[see e.g.][for more details]{{Mevius:2016ba}}.
Fig.~\ref{fig:panelIVsub} shows a panel of Stokes I and V images of the NCP with $\sim3'$ and $\sim$\eorFWHM$'$ FWHM resolution, after subtraction of the sky model. The Stokes-V images appear noise-like, whereas the Stokes-I images are classical confusion noise limited.

\paragraph{Diffuse Stokes Q \& U Emission}

In the power spectra analyses (Sect.~\ref{sec:powerspectra}) we {\sl only} use images made from  50-250$\lambda$ baselines as motivated in Sect.~\ref{sec:dataproc}. These short-baseline images indeed retain their diffuse Q and U power. Polarization leakage is assumed to be small \citep[see e.g.][]{Asad:2015vv, 2016arXiv160404534A}. In a forthcoming publication we will present the polarized structure of the NCP and its impact on the detection of the EoR signal in much deeper integrations.

\paragraph{Diffuse Stokes I Emission}

Diffuse Stokes-I emission is harder to detect when using 50-250$\lambda$ baselines, because it appears below the classical confusion noise level set by discrete sources. Images including the $10-50\lambda$ baselines clearly show diffuse emission, when averaged to lower resolution. Hence, the diffuse (EoR) emission should be retained in the images after \ddcal\ with \sagecalco.

\begin{figure*}[t]\label{fig:power-spec-2D-post-GMCA}
\centering
\includegraphics[width=0.9\hsize]{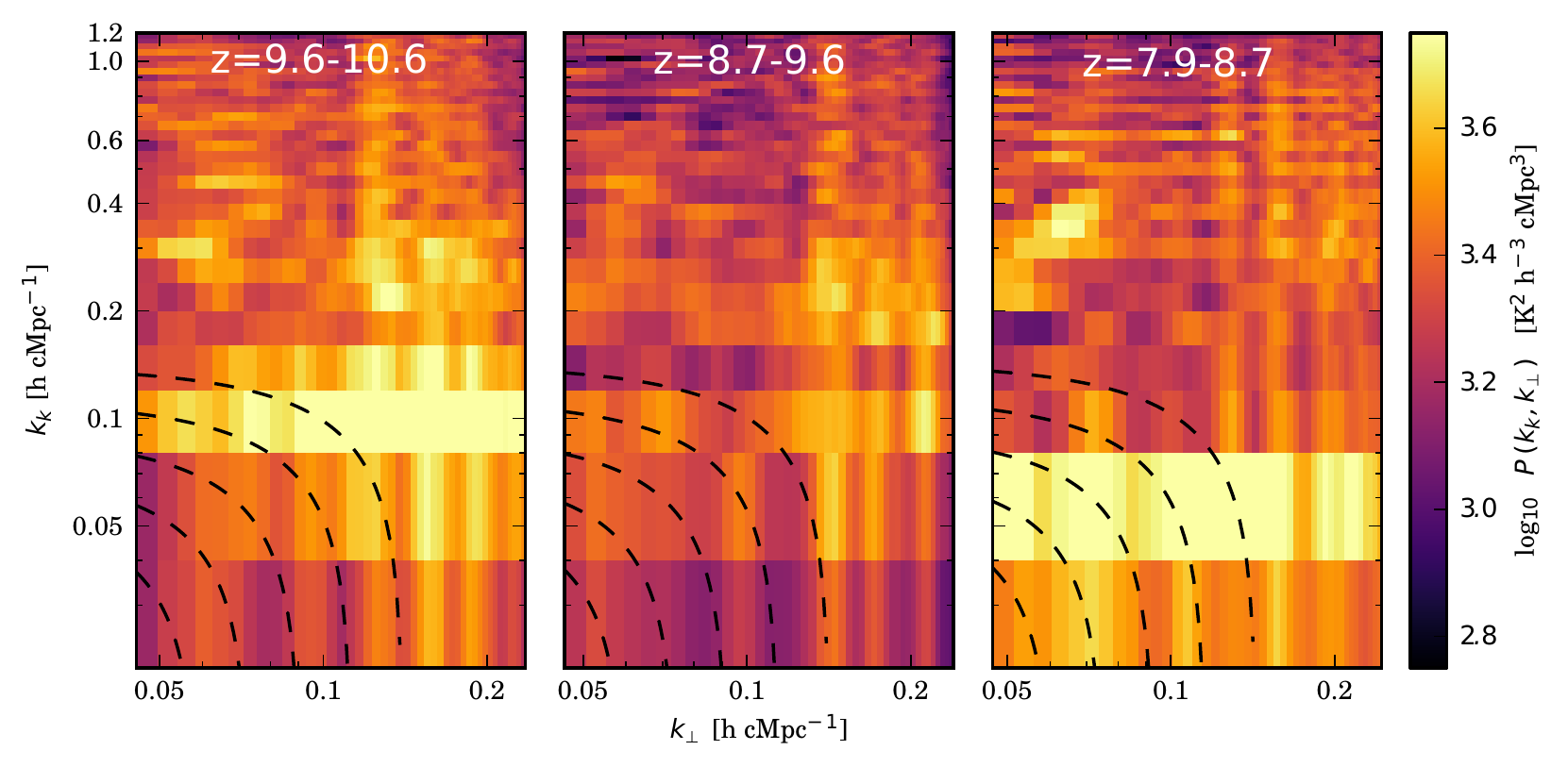}
\includegraphics[width=0.9\hsize]{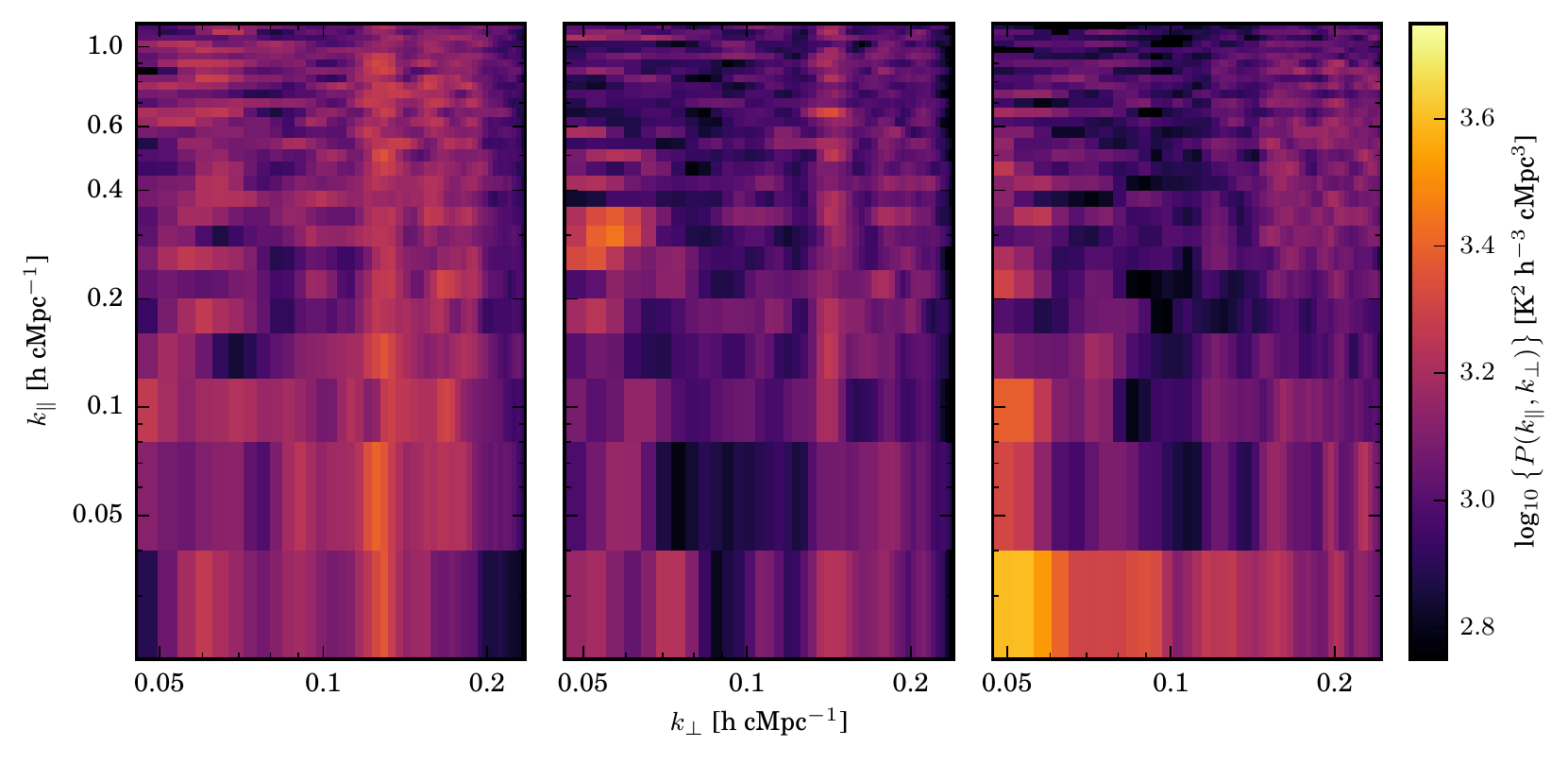}
\caption{Stokes I (top) and V (bottom) cylindrical power spectra after sky model and {\tt GMCA}-model subtraction, for \nightone. 
From left to right are shown the redshift ranges $z=$\,\zone, $z=$\,\ztwo\ and $z=$\,\zthree, respectively. The dashed curved lines in the Stokes I spectra refer to $k$ values of 0.054, 0.067, 0.083, 0.103 and 0.128 for $z=$\,\ztwo\ and only slightly different values for the other redshift bins. It is along these lines that we form the spherically averaged power spectra.}
\end{figure*}

\subsection{Generalized Morphological Component Analysis}\label{sec:GMCA}

The remaining foreground emission inside the primary beam area (Fig.\,\ref{fig:panelIVsub}) should only change very slowly with frequency and thus separable from the spectrally fluctuating 21-cm EoR signal \citep[e.g.][]{Morales:2004p3533}.  We use {\tt GMCA} \citep[][]{bobin07a, bobin07b, bobin07c, bobin08, bobin13}, specifically tailored to foreground removal \citep[][]{{2013MNRAS.429..165C}}, to remove the dominant modes from the data-cubes in Stokes I and any remaining instrumental polarization leakage in Stokes V.

GMCA is a blind source separation technique introduced by \citet[]{Zibulevsky:2001gw}, which uses as few assumptions about the data as possible in order to form a model of the foregrounds. The method works on the premise that the diffuse foregrounds consist of a number of statistically independent components which can be separated using the morphology of those components. An appropriate decomposition basis is sought such that the components appear sparse and in this analysis we use a wavelet decomposition. A component can then be easily separated from the other components, the cosmological signal and instrumental noise due to the components having only few significant basis coefficients which are likely to be different between components. This results in a foreground model which can be subtracted from the total data, leaving the sub-dominant cosmological signal and instrumental noise. The only user input to the default method is the number of components in the foreground model. The optimal choice for this could be led by a Bayesian model selection; however, previous analyses have shown that the foreground model is fairly robust to this choice \citep[see][for details]{{2013MNRAS.429..165C}} and as such we vary this number only over a limited range in this paper.

The implementation of {\tt GMCA} is the same as described in \citet{{2013MNRAS.429..165C}}. No astrophysical prior information is included in the calculation. While it is possible to include spectral information about the  foregrounds within the mixing matrix, we choose to implement {\tt GMCA} in the blindest way possible while the data is in the early stages of being constrained. The mixing matrix does not vary across the sky or across the wavelet scales as in more recent implementations \citep{Bobin:2013ci}. It is possible that the variation of the mixing matrix with wavelet scale may be implemented in a later data analysis as a method of mitigating the frequency-dependent PSF. Here we instead have chosen to set our data to a common resolution through $uv$-cuts in the imaging step and careful weighting. The solutions are regularised following Eqn.13 in \cite{Bobin:2013ci}, using $N_s$ components. The $p=0$ formalism is not trivial to calculate and the norm is relaxed to an L$^1$-norm with $p=1$, most often the standard in {\tt GMCA} implementations.

We note that {\tt GMCA} does not remove most of the remaining side-lobe noise. We remove $N_s$=6--8 components in Stokes I and $N_s$=0--2 components in Stokes V. The number of components are chosen to obtain an approximately flat noise behaviour in the $k_\parallel$ direction (see Table~\ref{tab:calsetup} for the exact numbers per redshift range). Fig.~\ref{fig:freqslice} shows a spatial-frequency slice through the Stokes-I data-cube after subtraction of the sky model. There are still spectrally smooth sources left in the data. After applying {\tt GMCA}, however, the Stokes-I data cube appears noise-like.
Finally, we note that whereas GMCA does not a-priori distinguishes foregrounds from the 21-cm EoR signal, extensive simulations by \citet[][]{chapman13} have shown that the 21-cm power-spectrum in the current range of $k$-modes should not be affected significantly by the diffuse and spectrally-smooth foreground removal.


\section{power spectra}\label{sec:results}

In this section we present the cylindrically and spherically averaged 21-cm power spectra. Using the former one can assess remaining systematics due to e.g.\ foreground residuals, side-lobe noise and frequency-coherent effects \citep[e.g.][]{Bowman:2009je, 2012ApJ...745..176V}. The latter achieves the highest signal-to-noise per $k$-mode. Given the relatively narrow LOFAR-HBA primary beam \citep[4.8--3.5$^\circ$ at 120--160\,MHz;][]{2013A&A...556A...2V} and our $3^\circ \times 3^\circ$ analysis window, we can ignore sky curvature. We use the Stokes-I residual data cube, after {\tt GMCA} (see e.g.\ Fig.\,\ref{fig:freqslice}), to measure the power spectra following \citet[][]{Tegmark:1997bm}. We use large enough cells that they can be assumed to be uncorrelated. 

\subsection{Power-spectrum Determination}\label{sec:powerspectra}

We first transform the data cube into brightness temperature, in units of mK \citep[see][for details]{08cdff67efba4dd6a63ff4635d4e10ad}. A Gaussian primary beam correction is applied, which is a good approximation over the $3^\circ \times 3^\circ$ analysis window \citep{haarlem13}, being smaller than the FWHM of the beam (see Figs.\,\ref{fig:FieldImage} \& \ref{fig:panelIVsub}). 
We account for $uv$-density weighting and the number of zero-valued $uv$-cells in the padded $uv$-grid\footnote{Due to the usual Jy\,PSF$^{-1}$ convention in radio astronomy, imagers scale $uv$-visibilities such that the zero-value visibility grid-cells are properly accounted for. The scaling, however, needs to be undone when determining the power spectrum.} that is used to create the image data cubes. Although determining the power spectrum directly from the (ungridded) visibilities is preferable, the size of the data set of 50\,Tbyte (Table~\ref{tab:dataproc}) renders this currently not feasible\footnote{Although the maximum information is retained in the un-gridded visibilities, gridding on scales substantially smaller than the inverse of the station beam ($\sim$$16\,\lambda$) -- in our case $4.58\,\lambda$ in the uv-domain (see Table~\ref{tab:calsetup}) -- should retain nearly all information.}. 
A second reason why we do not use the visibilities is that {\tt GMCA} is applied to the image cubes and not to the visibilities The inference of the power spectrum  follows \cite{Tegmark:1997bm, 2016arXiv160102073T} in part, but is adapted to the analysis of the image cube. 

To determine the power spectrum, we spatially Fourier transform the cube back to the $uv$-domain, and use a Least Squares Spectral Analysis (LSSA) method to transform the frequency axis into a delay axis ($\nu \leftrightarrow\tau$) \citep[see][]{Barning1963, Lomb1976, Stoica2009, 2016arXiv160102073T}, properly accounting for the missing channels due to RFI excision (see Fig.\,\ref{fig:freqslice} for the flagged channels). 

We transform all axes into inverse co-moving Mpc 
\citep[e.g.][]{{Morales:2004p3533}}, using the cosmological convention of $k=2\pi/L$. We determine power spectra $P(k)$ in units of K$^2$\,$h^{-3}$\,cMpc$^3$ or $\Delta^2(k)=k^3/(2 \pi^2) P(k)$ in units of K$^2$. We also use mK units, where more conventional.
Both the cylindrical and spherical power spectra are optimally weighted using the Stokes-V variance, down-weighting high noise-variance data \citep[e.g.][]{Tegmark:1997bm}.

\subsection{Cylindrical power spectra }

We present the power spectra for all redshift bins ($z=$\,\zthree, \ztwo\ and \zone, respectively) in Fig.\,\ref{fig:power-spec-2D-post-GMCA}, for both Stokes I (left) and Stokes V (right). We note the following:

\begin{itemize}

\item There is some banded structure in $k_\perp$ due to LOFAR-HBA $uv$-density variations modulating the noise variance in the Stokes-V power spectrum. No obvious structures in $k_\parallel$ are seen (e.g.\ "wedge"; \citet{Bowman:2009je, 2012ApJ...745..176V}). Before {\tt GMCA} polarization leakage appears in Stokes V in the lowest $k_\parallel$ bin, because of its broad-band nature. Because polarization leakage is also expected to be broad-band \cite[see e.g.][]{{Asad:2015vv}}, {\tt GMCA} effectively removes it with at most two components \citep[see][for a description of {\tt GMCA} components]{chapman13}. 

\item The Stokes-I power spectrum appears similar to that of Stokes V after {\tt GMCA} except for a residual horizontal band at $k_\parallel \approx 0.1$\,\cmpc in the $z=$\zone\ redshift bin and there is higher power in the $z=$\zthree\, redshift bin around $k_\parallel \approx 0.05$\,\cmpc. These are possibly caused by low-frequency structure remaining after the foreground removal with {\tt GMCA}. There is at most only a mild indication in Fig.\,\ref{fig:power-spec-2D-post-GMCA} for a wedge-like structure, suggesting that sky-model subtraction has been very effective, including the removal of side-lobes of out-of-beam sources.    

\item The ratios between the Stokes-I and Stokes-V power spectra for the three redshift bins is typically 2--3 in variance (see Fig.\,\ref{fig:ratioPS}).  Apart from the horizontal band at $k_\parallel \approx 0.1$\,\cmpc, in the $z=$\zone\ and a similar band at $k_\parallel \approx 0.05$\,\cmpc, at $z=$\zthree\,these plots are  devoid of significant features. The vertical bands have largely disappeared -- in agreement with the cause of the modulation arising as a result of variations in the $uv$-density. It also suggests that the \exv\ does not add coherently \citep[see also Fig. 10 in][]{Patil:2016td}, otherwise it would not average down with the number of visibilities in the same way as thermal noise which dominates Stokes V. No evidence for signals related to cable-reflections, at their known delays (or $k_\parallel$ values), is seen.  

\end{itemize}

We assume that the \exv\ is not the 21-cm EoR signal. It might be a mixture of side-lobe noise due to an incomplete and inaccurate sky model (Sect.~\ref{sec:calibration}) -- causing calibration gain errors \citep[e.g.][]{{2016MNRAS.461.3135B}} --, or effects of thermal and ionospheric noise, and {\sl leverage} \citep[e.g.][]{{Patil:2016td}}. We note that the \exn\ decreases as the gain solutions are regularised in the frequency direction (see Sect.\ref{sec:dataproc}). Because we split our baselines between calibration and imaging, and only subtract sources, but do not correct the residual visibilities after {\sl DI}-calibration, any suppression or enhancement of Stokes-I power must have its cause in the applications of the gains to the sky model. Hence they have to come from issues relevant for the longer baselines and the most likely effects are either an incomplete/inaccurate sky model or strong ionospheric variations. However, we have not seen evidence yet for correlations between the diffractive scale of the ionosphere and \exn\ in other data sets \citep[see Fig. 10 in][]{Patil:2016td}.  

\begin{figure*}[t]\label{fig:DDbeforeafter}
\centering
\includegraphics[width=0.9\hsize]{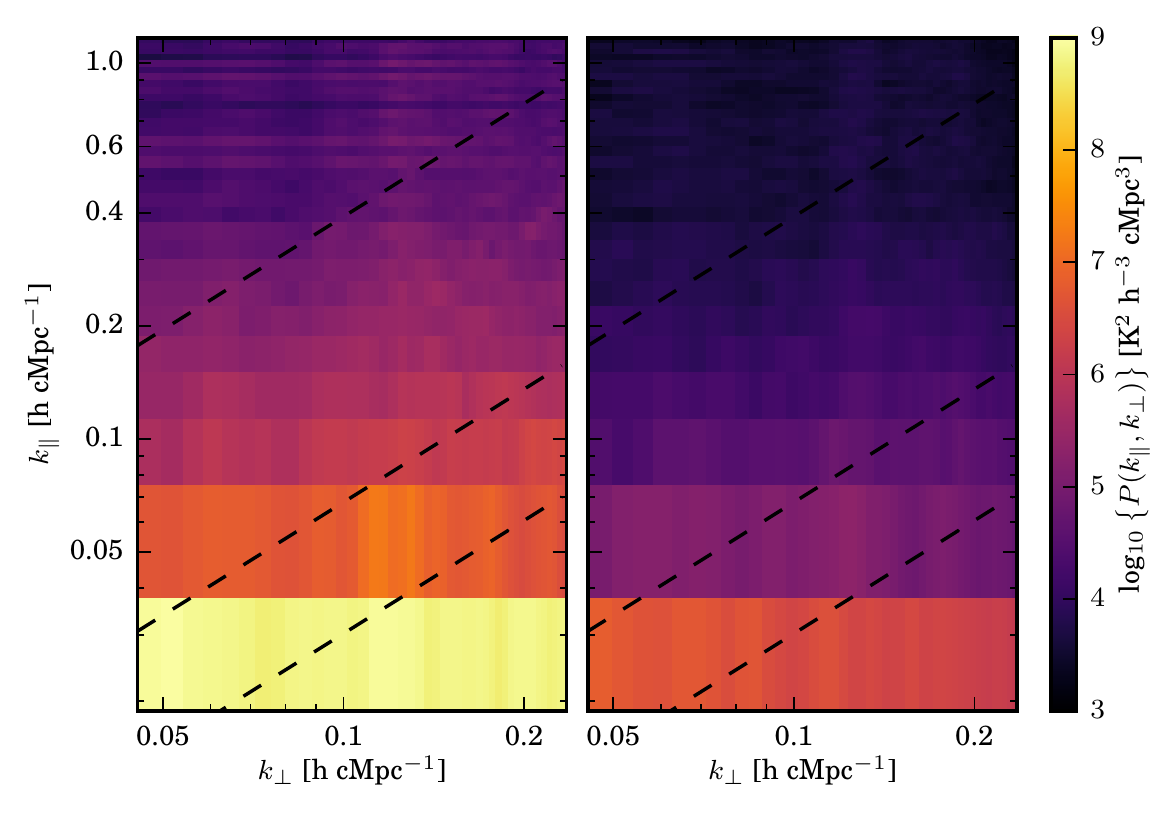}
\caption{The Stokes-I power spectra for the redshift range $z=$\,\zone, before (top left) and after (top right) \ddcal\ with \sagecalco, respectively.  Note the large drop in power of the foregrounds at low $k_\parallel$ and the removal of substantial power above the wedge as well. The dashed slanted lines indicate, from bottom to top, the location of angular distances of 4.5$^\circ$ and 10$^\circ$ from the phase centre, and the maximum delay corresponding to the horizon as seen from the zenith. The ratio between these power spectra is shown in Fig.~\ref{fig:DIDDratio}}
\end{figure*}

\begin{figure}[t]\label{fig:DIDDratio}
\centering
\includegraphics[width=\hsize]{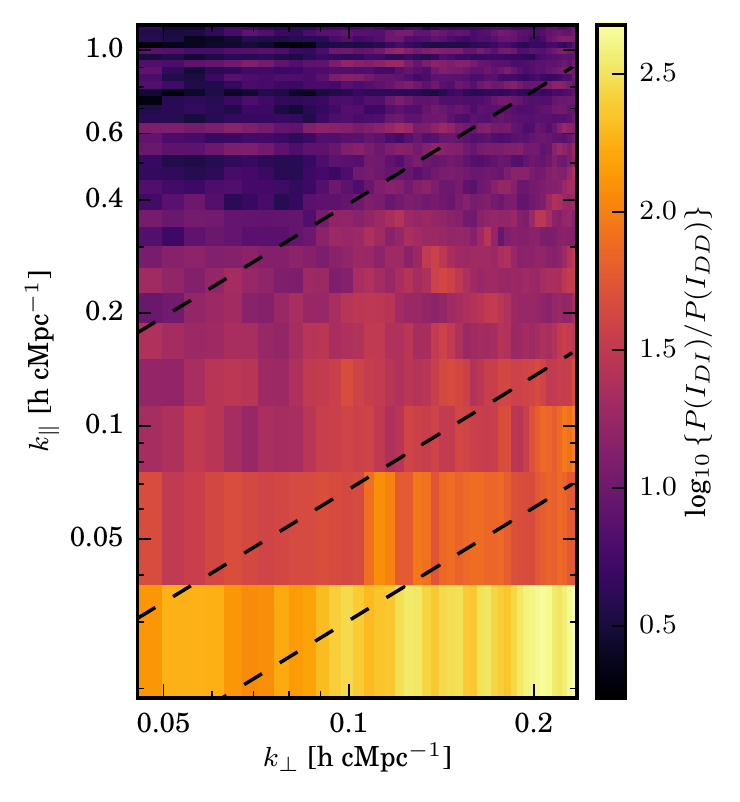}
\caption{The ratio between the Stokes-I power before and after \ddcal\ There is a drop of two orders of magnitude in power in the foregrounds at low $k_\parallel$.  The dashed slanted lines indicate, from bottom to top, the location of angular distances of 4.5$^\circ$ and 10$^\circ$ from the phase center, and the maximum delay corresponding to the horizon as seen from the zenith.}
\end{figure}

\begin{figure*}
\centering
\includegraphics[width=0.95 \hsize]{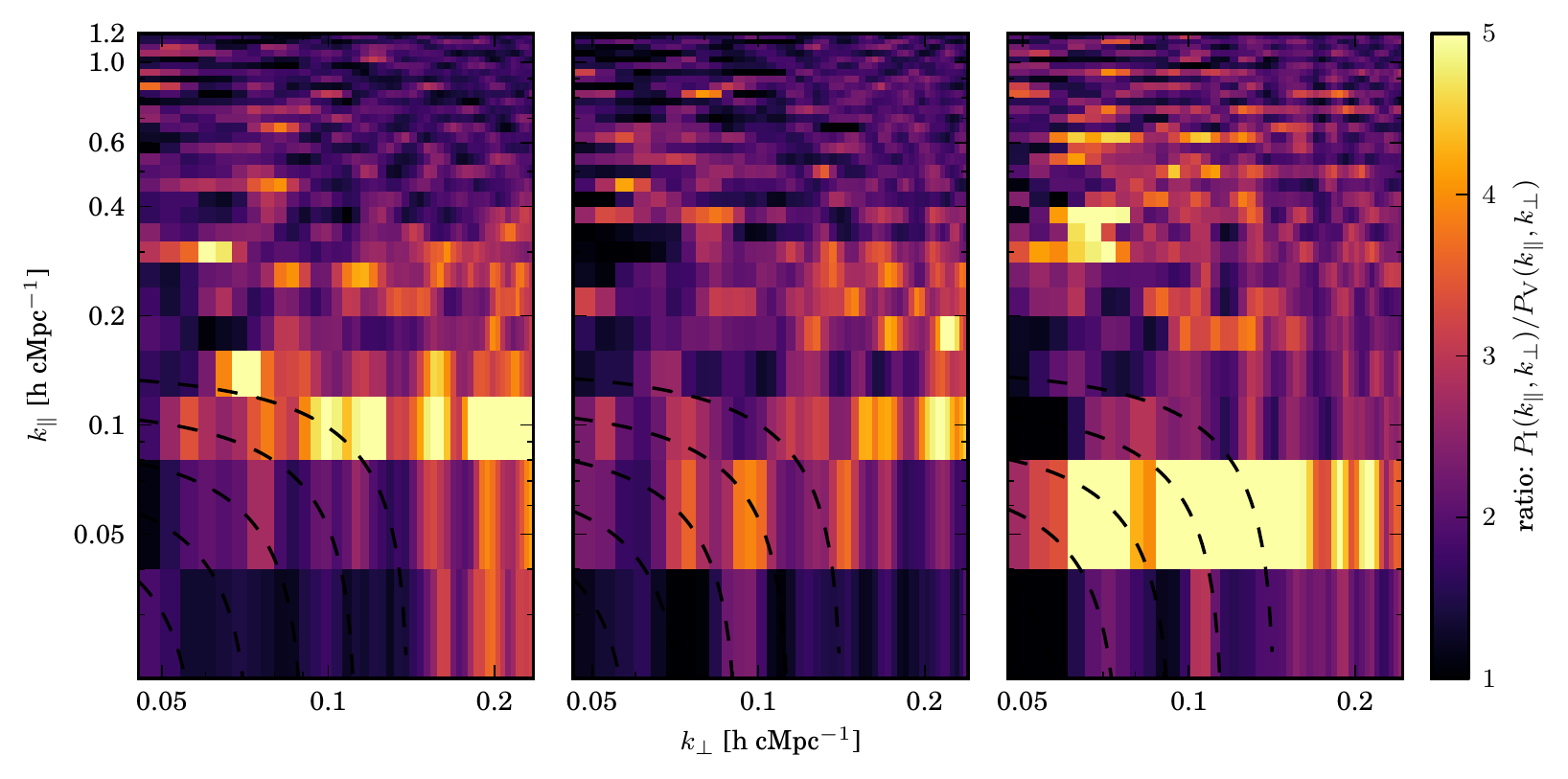}
\caption{{The Stokes I over V power spectra ratios for the redshift ranges $z=$\,\zone, $z=$\,\ztwo\ and $z=$\,\zthree, respectively.}}
\end{figure*}\label{fig:ratioPS}

To illustrate the considerable impact of \ddcal, we show the cylindrical power spectra for $z$=\,\zone\ before and after \ddcal\ and sky-model subtraction in Fig.~\ref{fig:DDbeforeafter}, and their ratio in Fig.~~\ref{fig:DIDDratio}, but before removal of the diffuse emission and residual sources in the primary beam with {\tt GMCA} (Sect.~\ref{sec:GMCA}).

\subsection{Spherical Power Spectra}

Next we determine the spherically-averaged power spectrum,  optimally weighting using the Stokes-V variance, following \cite{Tegmark:1997bm, 2016arXiv160102073T}, to obtain the average per $k$-bin. We flag two $k_\parallel$ bins with show strong \exv\  after running $\rm GMCA$ (see Fig.~\ref{fig:ratioPS}). In the $z=$\,\zone\ redshift range this corresponds to the (logarithmic) bin around $k_\parallel\sim 0.05$\,\cmpc. In the $z=$\,\zthree\ redshift range this corresponds to the bin around $k_\parallel\sim 0.1$\,\cmpc.  The integration is done along the curved lines shown in Fig.~\ref{fig:power-spec-2D-post-GMCA}.
We emphasize that we assume that the Stokes-V power-spectrum to be our best estimator of the thermal-noise power spectrum, because (i) the Stokes-V sky is by any means empty, and (ii) the thermal noise in Stokes V and I should be identical. Hence $\Delta_I^2-\Delta_V^2$ is the noise-bias corrected residual Stokes-I power spectrum. This should in principle be consistent with the 21-cm EoR power spectrum if there were {\sl no} excess variance nor other biases. Given the 13\,hr integration, however, this should still be considered an upper limit on the 21-cm EoR signal. We therefore conservatively put our upper limits at 2-$\sigma$ on top of the excess variance and do not attempt to estimate the excess variance level itself or correct for it at present (since we have no independent  estimator for it).

The resulting Stokes I, V and difference power spectra are shown in Fig.\,\ref{fig:powerspectra}, up to $k=0.2$\,\cmpc. The errors on the power spectra are determined from the Stokes-V variance and the number of {\sl uv} cells used in the integration. The errors are therefore plotted on the noise-bias-corrected powers. We note the following:

\begin{itemize}

\item The redshift ranges \zone\ and \ztwo\ appear power-law like\footnote{We note that such behaviour is only an approximation that would hold if $P(k)$ is roughly constant.} in the spherically averaged power spectrum (Fig.~\ref{fig:powerspectra}). Apart from two stripes, they also have mostly featureless ratios of Stokes-I over Stokes-V power (Fig.~\ref{fig:ratioPS}), 

\item Whereas at all $k$-values the Stokes I variance exceeds the Stokes V variance, given that the EoR signal very likely is still lower than the thermal noise, we have to assume that this \exv\ is due to other causes. We interpret it as a robust upper limit on the 21-cm emission power spectrum $\Delta^2_{21}$.

\item Up to $k_\perp \approx 0.2$\,\cmpc\ both the Stokes I and Stokes V power spectra follow approximate power-laws, with the power in Stokes I exceeding that in Stokes V for all $k$-modes and all redshift bins. At the smallest $k=\kbest$\,\cmpc, however, these values start to approach each other with only marginal differences. This is the bin that we regard as the best upper limit in terms of mK$^2$ sensitivity yielding a 2-$\sigma$ upper limit of $\Delta^2_{\rm 21} < (\ulbest{\rm ~mK})^2$ on the 21-cm power spectrum in the range $z=$\,\zbest.  

\end{itemize}

In Table~\ref{tab:upperlimits} we summarize the 2-$\sigma$ upper limits for the three redshift bins for $\Delta^2_{21}$. 


\section{Summary and Future Outlook}\label{sec:conclusions}

\begin{table}
\centering
\begin{tabular}{c c c c}
\hline
$k$ & $z$\,=\,\zthree & $z$\,=\,\ztwo & $z$\,=\,\zone \\
$h$ cMpc$^{-1}$ & mK$^2$ & mK$^2$ & mK$^2$\\
\hline
0.053 & (131.5)$^2$ & (86.4)$^2$ & (79.6)$^2$\\
0.067 & (242.1)$^2$ & (144.2)$^2$ & (108.8)$^2$\\
0.083 & (220.9)$^2$ & (184.7)$^2$ & (148.6)$^2$\\
0.103 & (337.4)$^2$ & (296.1)$^2$ & (224.0)$^2$\\
0.128 & (407.7)$^2$ & (342.0)$^2$ & (366.1)$^2$\\
\hline
\end{tabular}
\caption{$\Delta^2_{21}$ upper limits at the 2-$\sigma$ level.}
\end{table}\label{tab:upperlimits}

We have presented the first upper limits on the 21-cm power spectrum ($\Delta^2_{21}$) from the Epoch of Reionization, obtained with LOFAR-HBA, using one night of good data quality obtained toward the NCP. Our main numerical results can be summarised as follows:

\begin{itemize}

\item An \exv\ is  detected in Stokes I for all $k$ modes and redshift ranges, leading to our best although still non-zero $\Delta^2_{\rm I} = (\exvarbest)^2$\,mK$^2$ (1-$\sigma$) at $k=\kbest$\,\cmpc in the redshift range \zbest. The \exv\ is seen over the entire cylindrical power spectrum range. It appears constant with no obvious outstanding features such as cable reflections. 

\item The most stringent 2-$\sigma$ upper limit of $\Delta^2_{\rm 21} < (\ulbest{\rm ~mK})^2$ on the 21-cm power spectrum is found at $k=\kbest$\,$h$\,cMpc$^{-1}$ in the range $z=$\,\zbest. For reference, in the absence of \exv\, we would have reached a 2-$\sigma$ upper limit $\Delta^2_{\rm 21} < (\ulbestnoise{\rm ~mK})^2$ for the same  $k$ and $z$ ranges.

\item In Table~\ref{tab:upperlimits} we summarize the 2-$\sigma$ upper limits for the three redshift bins for a range of $k$-modes. 

\end{itemize}

Currently the cause of the \exv\ is still unknown. Based on simulations \citep[see e.g.][]{Patil:2016td} and data-processing tests, in particular with improved sky models and regularised gain solutions \citep[][]{{Yatawatta:2016we}}, it is likely due to residual side-lobe noise seen on the calibration baselines (due to an incomplete/inaccurate sky model), which affects the gain solutions on shorter baselines, as well as {\sl leverage} \citep[][]{Patil:2016td}. Various test are under way to find the cause, or causes. 
 
\subsection{Comparison of results}

Comparing our deepest 2-$\sigma$ upper limit of  $\Delta^2_{\rm 21} < (\ulbest{\rm ~mK})^2$ at $k=\kbest$\,$h$\,cMpc$^{-1}$ and $z=$\,\zbest, to those published by the other three teams (see Sect.~\ref{sec:intro}) using the GMRT \citep[see][]{paciga13}, the MWA \citep[see][]{Beardsley:2016tp} and PAPER \citep[see][]{ali15}, remains difficult. The reasons are the different redshift ranges and $k$-modes that are being quoted, as well as the considerably different integration times,
being 13\,h for LOFAR, 32\,h for MWA, 40\,h for the GMRT, and 1150\,h for PAPER, respectively, as well as  the use of very different instrumental configurations and post-correlation processing methods.

Currently, LOFAR-HBA reaches the highest redshift range of these experiments, with its deepest upper limits at $z$\,=\,\zone\ and only mildly less deep at $z$\,=\,\ztwo\, (Table~\ref{tab:upperlimits}). It also reaches considerably larger co-moving scales (i.e.\ smaller $k$-modes) compared to all other experiment, largely thanks to a strong emphasis on removal of compact sources and diffuse foreground emission from the data, allowing us to probe into the wedge region. 

\subsection{Lessons Learned}

We have learned that a number of requirements are important in the analysis of the LOFAR-HBA EoR data (see Sections~\ref{sec:dataproc} and \ref{sec:calibration}). We expect this to hold for other arrays as well \citep[see e.g.][for earlier discussions about the SKA]{2013ExA....36..235M}. Not meeting some of these requirements appears detrimental to our calibration and image quality (Sect.~\ref{sec:calibration} and \ref{sec:dataproc}), and the resulting power spectra:

\paragraph{\sl Direction-dependent calibration:} We use \dddir\ directions, clustering sources typically in (few) degree-scale patches (see Sect.~\ref{sec:calibration}). This scale roughly matches that expected based on the beam forming and isoplanatic angles, but are ultimately limited in size by the signal-to-noise per baseline and the number of degrees of freedom. 

\paragraph{\sl Completeness and accuracy of the sky model for calibration:} We use $\sim$\skymodelsources\ source components spread over about 19$^\circ$ in radius from the NCP (and beyond) down to flux-density levels of $\sim$\fluxlim\, mJy (inside/outside primary beam), below the classical confusion noise on short baselines (Sect.~\ref{sec:skymodel}). Our model does not yet include diffuse emission, especially the ubiquitous diffuse  polarized emission.

\paragraph{\sl Diffuse-emission conservation on the short baselines:} We currently use two non-overlapping baseline sets split at 250$\lambda$ (Sect.~\ref{sec:calibration}). Long baselines are used for calibration and short baselines for the power spectrum analyses. The fundamental reason is that  \ddcal\ suppresses diffuses emission in Stokes Q and U, and likely also the 21-cm EoR signal in Stokes I (Sect.~\ref{sec:suppression}).

\paragraph{\sl Wide-frequency domain for calibration:} To reduce the effects of \exn\ or \exv, due to leverage, side-lobe noise, ionospheric and thermal noise, etc., highly irregular gain solutions need to be penalised if not warranted by the data (Sect.~\ref{sec:calibration}). We have implemented this via regularisation of the gain solutions, using third-order Bernstein polynomials. \\

As noted in Sect.~\ref{sec:calibration}, \ddcal\ is necessary, but removes diffuse emission on short baselines, which is not part of the calibration model due to computational limits. Hence splitting the baselines in two sets (short and long) is necessary because diffuse emission is not measured on the longer (calibration) baselines (to our levels of sensitivity). This however leads to \exn, which is partly mitigates by using a larger frequency domain for the gain solutions. 

\subsection{Future Outlook}\label{sec:future}

Although the \exn\ has not yet been fully eliminated, gain regularisation over a large frequency domain, as implemented in {\sl SageCal-CO} \citep[][]{{Yatawatta:2016we}}, has considerably reduced its magnitude in recent analyses. To reduce the \exv\ further, by a factor 2--3, i.e.\ to the level approaching Stokes-V power on all $k$-modes, we plan to:

\begin{itemize}

\item Improve the calibration sky model by including even fainter compact sources inside and outside the primary beam. With the current \skymodelsources\ component model, we still notice improvements when new sources are added.

\item Include diffuse Stokes Q, U and (if possible) diffuse Stokes I emission in the sky model and, if possible, avoid the split-baseline
approach. This should reduce the \exv\ as tests have shown, due to the elimination of {\sl leverage}, while not suppressing diffuse emission.

\item Improve {\tt GMCA} foreground subtraction, or replace it by a spectrally-smooth diffuse foreground model and subtract it in the $uv$-plane on short baselines.

\item Use the cross-variance between different observing epochs and assess whether the \exv\ is (in)coherent. This approach avoids the need for a careful noise-power estimate and its bias correction in the Stokes-I power spectrum. 

\item Cross-correlate the gain solutions with data-quality metrics (e.g.\ diffractive scale) and sky- and calibration-model metrics to gain better insight into the nature of the \exv.

\item Include the flagged interferometers between co-located stations sharing the same electronics cabinet -- with baselines in the range of 40--60$\lambda$ -- in the analysis. Although these baselines are the most sensitive to the 21-cm signal, they were conservatively flagged to avoid any correlated spurious signals. We have started a program to statistically analyse the signals on those baselines to quantify any non-celestial contributions and include as many of them as possible.

\item Analyse the full set of data, in steps, and combine their results. If the \exv\ is incoherent, and if all nights turn out to be of similar quality, we should be able to reduce the upper limits inverse proportional with integration time (in power spectrum variance). From an earlier analysis of several nights we have indications that the excess noise is indeed only weakly correlated between nights \citep[see][]{Patil:2016td}.

\end{itemize}

The results presented in this paper show that the LOFAR residual images and power spectra are still affected by low-level effects  (e.g.\ \exv). However, we have identified viable mitigation strategies to reduce its level. Given that the results in this paper are (i) based on only $\sim$2\% of the entire NCP data set in hand and (ii) still conservatively excludes some of the most sensitive short baselines, we are confident that we can reach considerably deeper limits in the near future.

\begin{figure*}
\centering
\includegraphics[width=0.98\hsize]{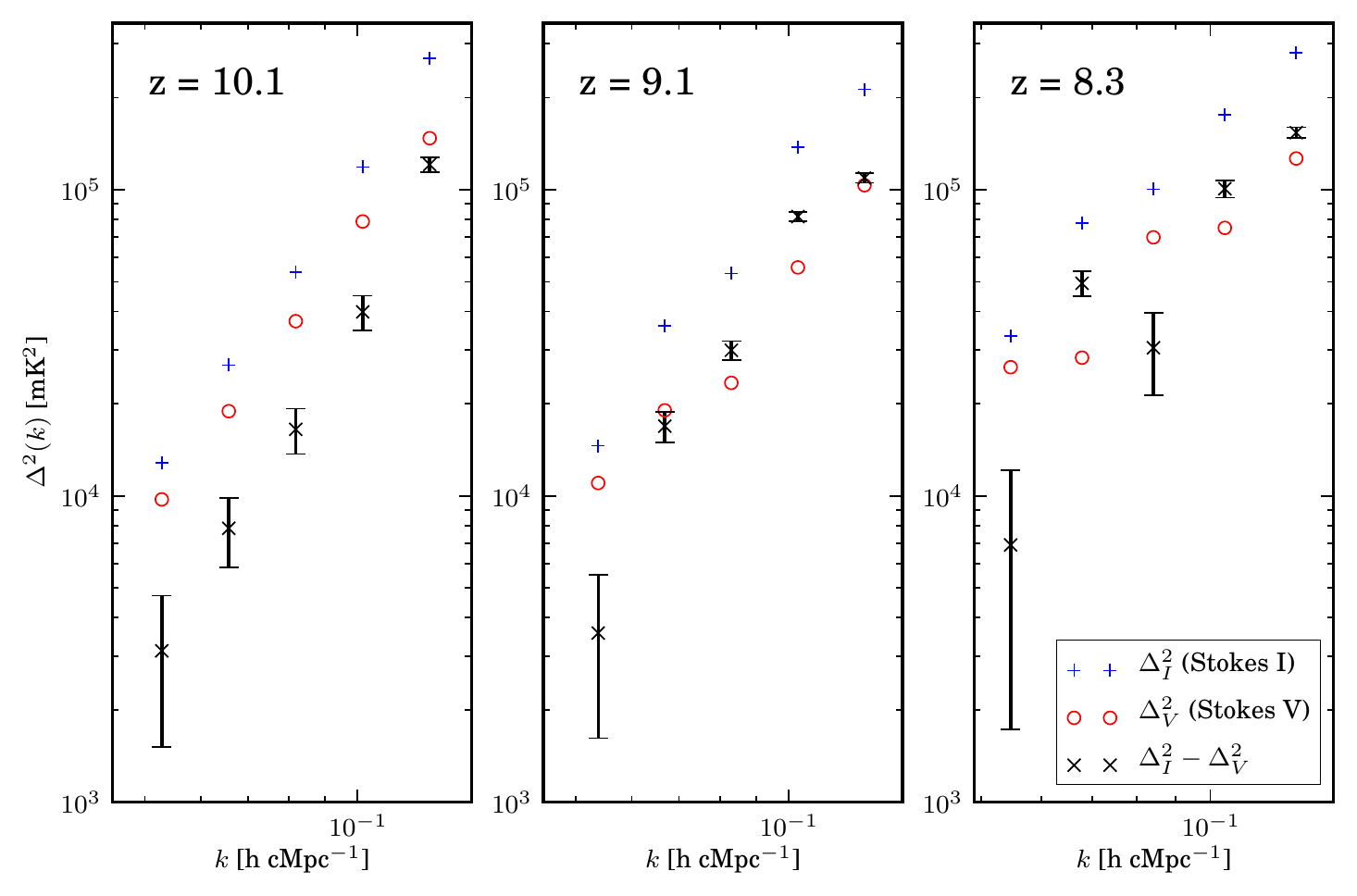}
\caption{{The spherically averaged Stokes I and V power spectra after GMCA for \nightone; From left to right are shown the redshift ranges $z=$\,\zone, $z=$\,\ztwo\ and $z=$\,\zthree\ from left to right, respectively. The mean redshifts are indicated in the panels.}}
\vspace{0.0cm}
\end{figure*}\label{fig:powerspectra}

\acknowledgments

LOFAR, the Low Frequency Array designed and constructed by ASTRON, has facilities in several countries, that are owned by various parties (each with their own funding sources), and that are collectively operated by the International LOFAR Telescope (ILT) foundation under a joint scientific policy.
SZ and AP would like to thank the Netherlands Organisation for Scientific Research (NWO) VICI grant for financial support. LVEK, HV, AG, SD, BKG thank the European Research Council Starting Grant (639.043.308) for support. AGdB, ARO, SBY, VNP, MM, HV and MH, acknowledge support from the ERC (grant 339743, LOFARCORE). MM, VNP and SBY also acknowledge support from the NWO TOP grant (614.001.005). VJ would like to thank the Netherlands Foundation for Scientific Research (NWO) for financial support through VENI grant 639.041.336. ITI was supported by the Science and Technology Facilities Council [grant number ST/I000976/1]. SM would like to acknowledge the financial assistance from the European Research Council under ERC grant number 638743-FIRSTDAWN and from the European Unions Seventh Framework Programme FP7-PEOPLE-2012-CIG grant number 321933-21ALPHA.  

\bibliographystyle{aasjournal}
\bibliography{main.bbl}

\end{document}